\documentclass[twoside,9pt]{article}
\usepackage{extsizes}
\usepackage[super,sort&compress,comma]{natbib} 
\usepackage[version=3]{mhchem}
\usepackage[left=1.5cm, right=1.5cm, top=1.785cm, bottom=2.0cm]{geometry}
\usepackage{balance}
\usepackage{mathptmx}
\usepackage{sectsty}
\usepackage{graphicx} 
\usepackage{lastpage}
\usepackage[format=plain,justification=justified,singlelinecheck=false,font={stretch=1.125,small,sf},labelfont=bf,labelsep=space]{caption}
\usepackage{float}
\usepackage{fancyhdr}
\usepackage{fnpos}
\usepackage[english]{babel}
\addto{\captionsenglish}{%
  
}
\usepackage{array}
\usepackage{droidsans}
\usepackage{charter}
\usepackage[T1]{fontenc}
\usepackage[usenames,dvipsnames]{xcolor}
\usepackage{setspace}
\usepackage[compact]{titlesec}
\usepackage{hyperref}
\usepackage{booktabs}
\PassOptionsToPackage{hyphens}{url}\usepackage{hyperref}

\definecolor{cream}{RGB}{222,217,201}

\newcommand{\calD}{\mathcal{D}}

\newcommand{\calM}{\mathcal{M}}

\newlength{\boxwidth}
\def\dd{\;\!\mathrm{d}}

\def\btheorem{\begin{theorem}}
\def\etheorem{\end{theorem}}
\def\blemma{\begin{lemma}}
\def\elemma{\end{lemma}}
\def\bproposition{\begin{proposition}}
\def\eproposition{\end{proposition}}
\def\bcorollary{\begin{corollary}}
\def\ecorollary{\end{corollary}}
\def\bdefinition{\begin{definition}}
\def\edefinition{\end{definition}}
\def\bexample{\begin{example}}
\def\eexample{\end{example}}
\def\bremark{\begin{remark}}
\def\eremark{\end{remark}}

\newcommand{\be}{\begin{equation}}
\newcommand{\ee}{\end{equation}}
\newcommand{\beq}{\begin{eqnarray}}
\newcommand{\eeq}{\end{eqnarray}}
\newcommand{\bem}{\begin{multline}}
\newcommand{\eem}{\end{multline}}
\newcommand{\ba}{\begin{align}}
\newcommand{\ea}{\end{align}}

\renewcommand{\figurename}{Figure}
\renewcommand{\tablename}{Table}

\begin{document}

\pagestyle{fancy}
\thispagestyle{plain}
\fancypagestyle{plain}{
\renewcommand{\headrulewidth}{0pt}
}

\makeFNbottom
\makeatletter
\renewcommand\LARGE{\@setfontsize\LARGE{15pt}{17}}
\renewcommand\Large{\@setfontsize\Large{12pt}{14}}
\renewcommand\large{\@setfontsize\large{10pt}{12}}
\renewcommand\footnotesize{\@setfontsize\footnotesize{7pt}{10}}
\makeatother

\renewcommand{\thefootnote}{\fnsymbol{footnote}}
\renewcommand\footnoterule{\vspace*{1pt}%
\color{cream}\hrule width 3.5in height 0.4pt \color{black}\vspace*{5pt}} 
\setcounter{secnumdepth}{5}

\makeatletter 
\renewcommand\@biblabel[1]{#1}            
\renewcommand\@makefntext[1]%
{\noindent\makebox[0pt][r]{\@thefnmark\,}#1}
\makeatother 
\renewcommand{\figurename}{{Fig.}~}
\sectionfont{\sffamily\Large}
\subsectionfont{\normalsize}
\subsubsectionfont{\bf}
\setstretch{1.125} %
\setlength{\skip\footins}{0.8cm}
\setlength{\footnotesep}{0.25cm}
\setlength{\jot}{10pt}
\titlespacing*{\section}{0pt}{4pt}{4pt}
\titlespacing*{\subsection}{0pt}{15pt}{1pt}

\fancyfoot{}
\renewcommand{\headrulewidth}{0pt} 
\renewcommand{\footrulewidth}{0pt}
\setlength{\arrayrulewidth}{1pt}
\setlength{\columnsep}{6.5mm}
\setlength\bibsep{1pt}

\makeatletter 
\newlength{\figrulesep} 
\setlength{\figrulesep}{0.5\textfloatsep} 

\newcommand{\topfigrule}{\vspace*{-1pt}%
\noindent{\color{cream}\rule[-\figrulesep]{\columnwidth}{1.5pt}} }

\newcommand{\botfigrule}{\vspace*{-2pt}%
\noindent{\color{cream}\rule[\figrulesep]{\columnwidth}{1.5pt}} }

\newcommand{\dblfigrule}{\vspace*{-1pt}%
\noindent{\color{cream}\rule[-\figrulesep]{\textwidth}{1.5pt}} }

\makeatother

  \begin{@twocolumnfalse}
\vspace{1em}
\sffamily
\begin{tabular}{m{4.5cm} p{12.85cm} }

 & \noindent\LARGE{\textbf{Deep learning reveals key predictors of thermal}} \\
& \noindent\LARGE{\textbf{conductivity in covalent organic frameworks$^\dag$}}\\
\vspace{0.3cm} & \vspace{0.3cm} \\

 & \noindent\large{Prakash Thakolkaran,$^{\ddag}$\textit{$^{a}$} Yiwen Zheng,$^{\ddag}$\textit{$^{b}$} Yaqi Guo,\textit{$^{a}$} Aniruddh Vashisth$^{\P}$\textit{$^{b}$}, } \\
 & \large{and Siddhant Kumar$^{\P}$\textit{$^{a}$} } 
\\

 & \noindent\normalsize{The thermal conductivity of covalent organic frameworks (COFs), an emerging class of nanoporous polymeric materials, is crucial for many applications, yet the link between their structure and thermal properties remains poorly understood. Analysis of a dataset containing over 2,400 COFs reveals that conventional features such as density, pore size, void fraction, and surface area do not reliably predict thermal conductivity. To address this, an attention-based machine learning model was trained, accurately predicting thermal conductivities even for structures outside the training set. The attention mechanism was then utilized to investigate the model's success. The analysis identified dangling molecular branches as a key predictor of thermal conductivity, {leading us to define the dangling mass ratio (DMR), a descriptor that quantifies the fraction of atomic mass in dangling branches relative to the total COF mass. Feature importance assessments on regression models confirm the significance of DMR in predicting thermal conductivity.}  These findings indicate that COFs with dangling functional groups exhibit lower thermal transfer capabilities. Molecular dynamics simulations support this observation, revealing significant mismatches in the vibrational density of states due to the presence of dangling branches.} \\

\end{tabular}

 \end{@twocolumnfalse} \vspace{0.6cm}

\renewcommand*\rmdefault{bch}\normalfont\upshape
\rmfamily
\section*{}
\vspace{-1cm}

\footnotetext{\textit{$^{a}$~Department of Materials Science and Engineering, Delft University of Technology, 2628 CD Delft, The Netherlands}}
\footnotetext{\textit{$^{b}$~Department of Mechanical Engineering, University of Washington, Seattle, WA, USA}}
\footnotetext{\dag~Supplementary Information available. See DOI: 00.0000/xxxxxxxx.}
\footnotetext{\ddag~These authors contributed equally to this work.}
\footnotetext{\P~These authors contributed equally to this work. E-mail: Vashisth@uw.edu ; E-mail: Sid.Kumar@tudelft.nl}

\section*{Introduction}

Covalent organic frameworks \cite{wang2020covalent,zhao2021covalent} (COFs) are an emerging class of nanoporous polymeric materials. Compared to metal-organic framework (MOFs) \cite{liu2021mof,stock2012synthesis} and zeolites \cite{perez2022zeolites}, the crystalline backbone of COFs is composed of organic building blocks -- known as knots and linkers -- connected by strong covalent bonds, thereby offering higher stability \cite{abuzeid2021covalent}.
The crystalline nature of COFs, along with their high surface areas, tunable pore sizes, and functionalizable organic linkers, make them exceptionally suited for a wide range of promising applications. These applications include (but are not limited to) photoconductivity \cite{dogru2013photoconductive, stegbauer2018tailor}, chemo-sensing \cite{ali2024thiazole, li2020fabrication}, catalysis \cite{yusran2020covalent}, drug delivery \cite{scicluna2020evolution,guo2022covalent}, thermoelectrics \cite{wang2017fluorene,chumakov2016first}, semiconductors \cite{wang2020semiconductive}, gas storage and separation \cite{fan2018covalent,wu2017applications}.

A key property dictating the applications of COFs is thermal transfer. For example, low thermal conductivity is desired for thermoelectrics to maintain large internal thermal gradients and increase efficiency \cite{zhou2018routes}. On the other hand, a high thermal conductivity is desired for gas adsorption and separation where efficient thermal transfer is important for the longevity and stability of the nanoporous membranes \cite{evans2021trends}. COFs are promising nanoporous candidates for such applications as the ability to design their crystalline topology (e.g., lattice type, pore size) and chemistry via choice of knots and linkers opens up a large and diverse space of thermal conductivities \cite{zou2013topology, pang2017regulating,gui2020three,wang2020covalent,zhao2021covalent}.  Therefore, it is of great interest to gain a deeper understanding of the thermal transfer mechanism and to obtain strong structure-property trends, which in turn enables the application-specific design of COFs.

Two approaches can be used to elucidate the thermal structure-property relationships of COFs: experimental trial-and-error methods which involve synthesis and characterization \cite{xu2018thermal, choy1977thermal}, and computational high-throughput screening which relies on first-principle calculations \cite{utimula2019ab, lindsay2012thermal, romero2015thermal}. With COFs offering practically an unlimited design space, an experimental  trial-and-error approach to screen new COF candidates (including developing new synthesis routes for each candidate) is prohibitively inefficient. Moreover, experimentally synthesized COFs will always contain crystalline defects \cite{li2021impact, li2020defective}, which greatly influence the thermal conductivity. This makes it difficult to relate the thermal transfer mechanisms to the geometrical and chemical make-up of a COF structure.  On the contrary, molecular dynamics (MD) simulations \cite{rapaport2004art} offer a high-throughput virtual screening alternative to lab-based experiments. However, despite recent advances in computing hardware, even virtual screening can be prohibitively inefficient for a large design space. This is highlighted by the example that generating a dataset of just 2,471 two-dimensional COFs and their thermal conductivity for this study required 1.3 million CPU-hours in cloud computing (translating into approximately four months and \$70,000 in time and cost, respectively). This underscores the need for efficient structure-property maps that are accurate, generalizable, and interpretable. 

\begin{figure*}[t]
\centering
\includegraphics[width=\linewidth]{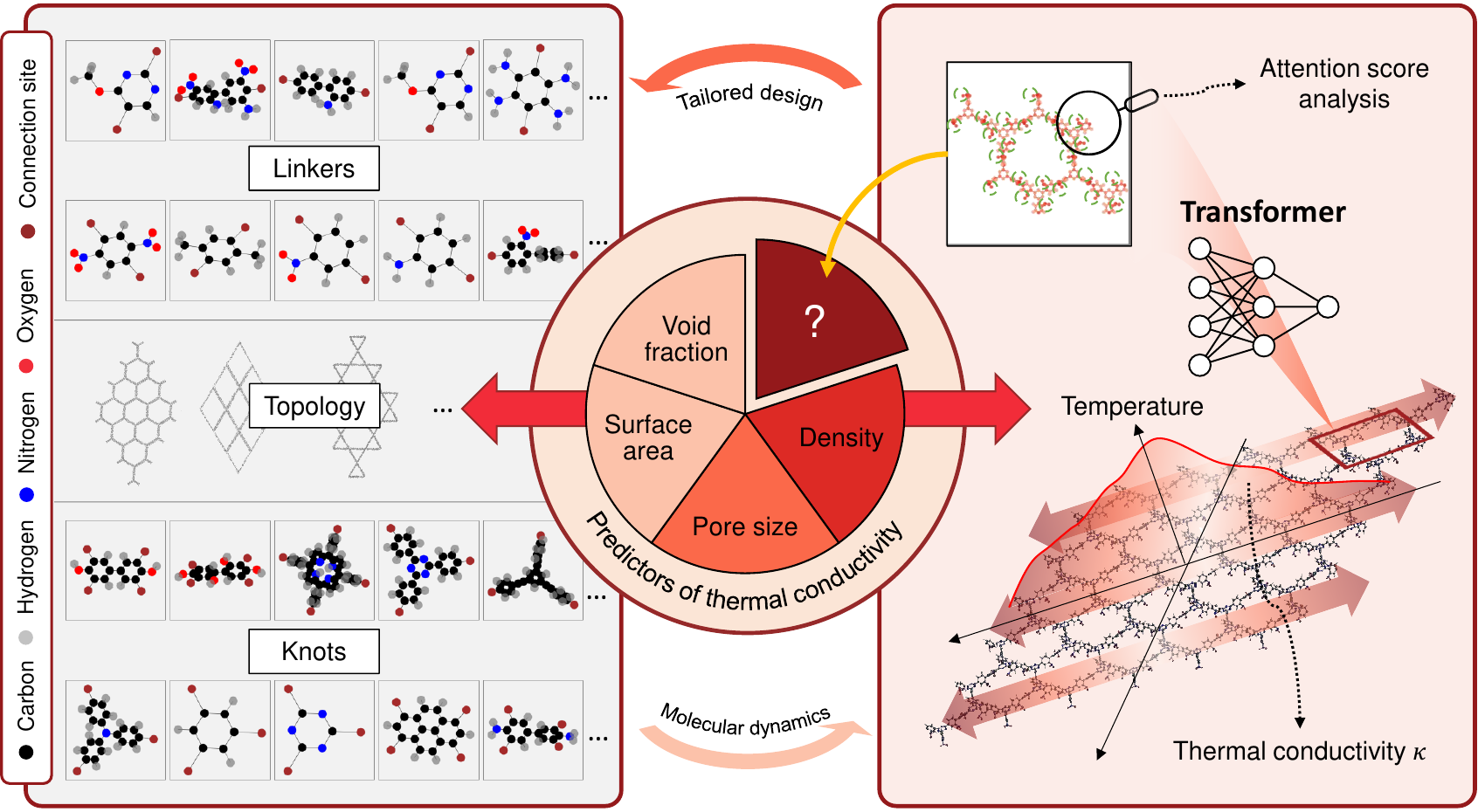}
\caption{Overview of this study. A COF structure consists of molecular substructures called linkers and knots arranged in a periodic pattern described by the topology. Through molecular dynamics, we construct a large dataset of 2,471 COFs (with diverse linkers, knots, and topologies) and their corresponding thermal conductivities.  We demonstrate that conventional descriptors, such as density, pore size, void fraction, and surface area, fail to predict the thermal conductivity reliably. We employ a  machine learning framework based on attention mechanism and transformer architecture to uncover a novel predictor, thus enhancing our understanding of the structure-property relationship of thermal conductivity of COFs.}
\label{fig:overview}
\end{figure*}

Recently, Islamov et al. \cite{islamov2023high} performed a high-throughput screening study of over 10,000 MOF structures and their thermal conductivity ($\kappa$) using MD. They found that the majority of the MOFs possess $\kappa < 1 \ \text{W}\text{m}^{-1}\text{K}^{-1}$ with a few exceptions possessing ultra-high thermal conductivity ($\kappa > 10 \ \text{W}\text{m}^{-1}\text{K}^{-1}$). While COFs are generally more thermally stable (owing to the strong covalent bonding of the building blocks), a recent study by Thakur et al. \cite{thakur2024pushing} on high-throughput screening study on over 10,000 COFs demonstrated that, generally, COFs also possess a similar range of $\kappa$, with the majority of the structures exhibiting $\kappa<1 \ \text{W}\text{m}^{-1}\text{K}^{-1}$. Interestingly, MOFs and COFs possess similar structure-property trends. \textit{(i)} Increasing the pore size is mostly sufficient to achieve a low thermal conductivity. \textit{(ii)} Additionally, one can increase the void fraction, the surface area, or include heavy atoms to further decrease thermal conductivity. \textit{(iii)} However, to increase the thermal conductivity, there are many factors that need to align. Islamov et al. \cite{islamov2023high} demonstrated that while lowering the pore size and increasing the density results in a higher ceiling of attainable thermal conductivity, other factors, such as the topology, mass-mismatch, and linker length also need to be accounted for. Thakur et al.  \cite{thakur2024pushing} came to similar conclusions alongside the observation that aligning the polymeric chains in the structure to heat flow direction raises the ceiling of attainable thermal conductivity. However, these hand-crafted guidelines and ad hoc correlations are insufficient for developing an accurate and generalizable predictive model that captures the thermal conductivity structure-property relationships of COFs. Our study shows that no combination of commonly used descriptors consistently predicts thermal conductivity. Consequently, a definitive method for identifying COFs with tailor-made thermal properties for various applications remains elusive.

To address this knowledge gap, we turn to machine learning (ML) with an emphasis on interpretability and explainability. {Previous efforts in ML-assisted design for thermally conductive organic materials include the discovery of polymers with high $\kappa$ through a hierarchical feature selection process \cite{huang2023exploring}, a reinforcement learning approach using SMILES-based representations \cite{ma2022exploring}, and methods leveraging molecular fingerprints as input features, such as fine-tuning a pre-trained regression model followed by screening \cite{wu2019machine}, training a neural network \cite{ishikiriyama2022polymer}, and employing an active learning approach \cite{xu2024unlocking}. Hu et al. \cite{hu2024thermally} provide a review of recent efforts and outlook on ML for thermally conductive organic materials.} For the ML modeling of structure-property maps of porous polymers (including but not limited to COFs) there are primarily two routes:
\begin{itemize}%
    \item \textit{high interpretability, limited accuracy:} using hand-crafted and pre-extracted high-level of the crystalline network (e.g., pore size, density, surface area, atomic composition)  as input to classical regression algorithms \cite{yang2021accelerating,cao2022machine,de2024high};
    \item \textit{high accuracy, limited interpretability:} using information-rich but raw graph representation of the crystalline network (where nodes denote atoms with features such as position and atom type, and edges denote bonds with the bond order as the feature) as input to end-to-end deep learning-based regression algorithms \cite{zhao2024pacman,korolev2024coarse,kang2023multi,park2023enhancing}.
\end{itemize} 
Here, we bridge the two routes to develop an accurate predictive model using deep learning while also combining interpretability insights from deep learning with prior knowledge and descriptors to explain the thermal conductivity structure-property relationships of COFs (see \figurename\ref{fig:overview} for an overview). 

In the following, we first conduct a large-scale data analysis of the thermal conductivity structure-property relations of COFs and identify the deficiencies in commonly used descriptors. Next, we provide a deep learning model that accurately predicts the thermal conductivity of COFs. Subsequently, an analysis {of the attention scores} of the deep learning model uncovers the presence of dangling atoms in the crystalline network of COFs as a strong and so-far missing key predictor for thermal conductivity. Additional physics-based analysis sheds light on the important role of dangling atoms in lowering the thermal conductivity of COFs by disrupting heat transfer pathways. We close by utilizing the ML model for efficient high-throughput screening and identifying COFs with extreme thermal conductivities.

\section*{Results}
\subsection*{High-throughput data analysis}

\begin{figure*}[t]
\centering
\includegraphics[width=\linewidth]{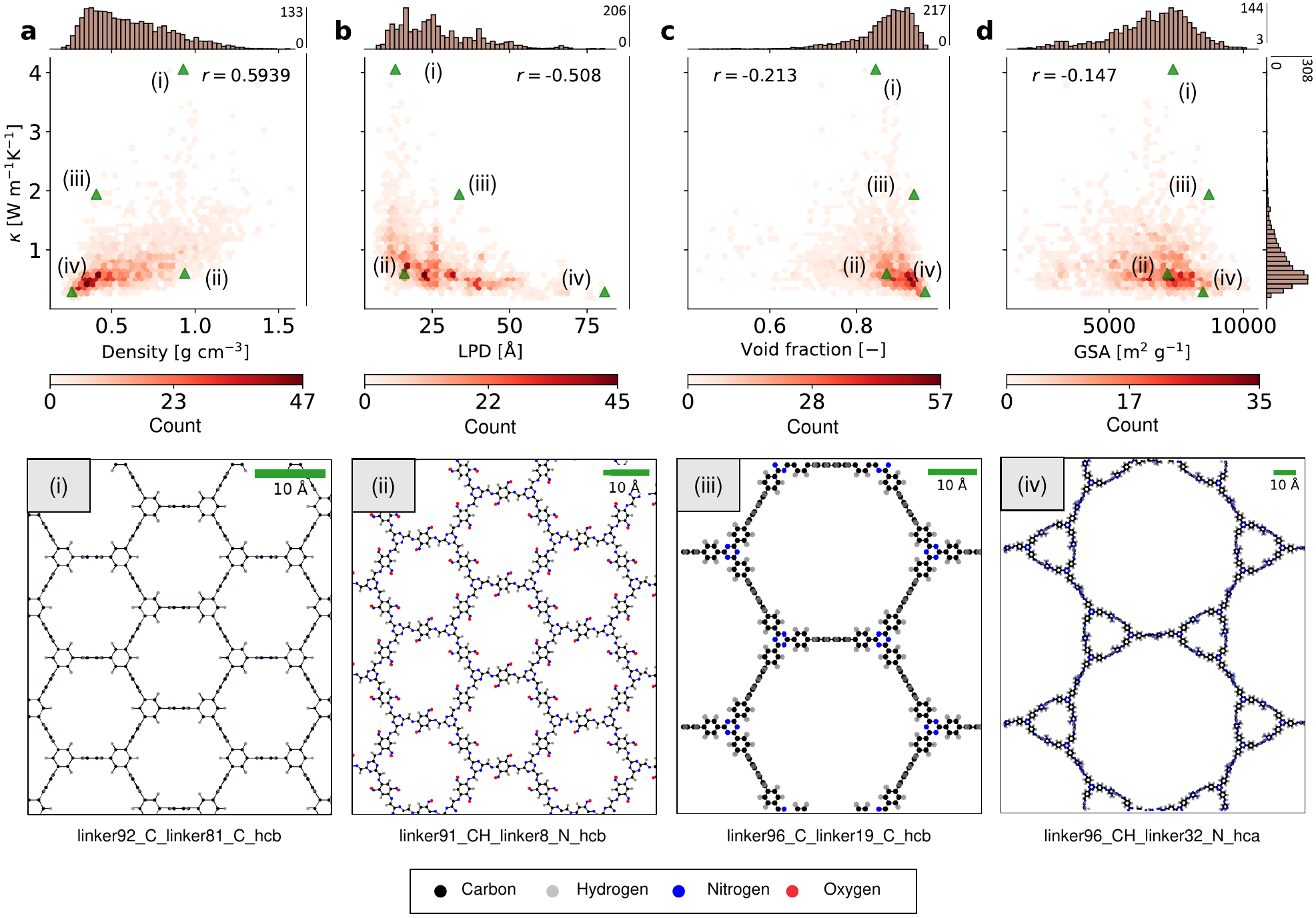}
\caption{Distribution of (\textbf{a}) $\kappa$ versus density  (with the color indicating data count per bin), (\textbf{b}) $\kappa$ versus largest pore diameter (LPD), (\textbf{c}) $\kappa$ versus void fraction, and (\textbf{d}) $\kappa$ versus gravimetric surface area (GSA). Also shown are four  COF structures (\textbf{i-iv}) with contrasting properties, marked by green triangles in the plots above. {The $r$-value indicates the Pearson correlation coefficient.}}
\label{fig:distribution}
\end{figure*}

We create a labelled dataset of COFs and their thermal conductivities by selecting 2,471 two-dimensional COFs from the unlabelled dataset of Mercado et al. \cite{mercado2018silico} The COF structures in this design space are made up of one type of nodal linker (also referred to as \textit{knot}) and one type of connecting linker. The sampled subset consists of COFs with 104 different linkers arranged in 25 different topologies (e.g., honeycomb, kagome, square lattice, etc.), and bonded by four different types of covalent linkages (i.e., carbon-carbon, amine, amide, and imine).  
We conduct non-equilibrium molecular dynamics (NEMD)  simulations to calculate thermal conductivities in two orthogonal in-plane directions (see Supplementary Information Section 1.1 for more details). 
{The resulting dataset shows a strong skew toward low thermal conductivity values, with the majority of COFs exhibiting $\kappa < 1 \ \mathrm{W m^{-1} K^{-1}}$.}
In Supplementary Information Section 1.2, we show that there is minimal in-plane anisotropy in thermal conductivity across the dataset. Therefore, we utilize average thermal conductivity in both directions, denoted as $\kappa$, as the quantity of interest to explore the structure-property relationships of COFs. In the following and in \figurename\ref{fig:distribution}, we present the observations from data analysis of the key descriptors of COFs and thermal conductivities in our labelled dataset. The distribution of all features and their correlations are detailed in Section 1.3 of the Supplementary Information.

\textbf{Density:} \figurename\ref{fig:distribution}a illustrates the distribution of $\kappa$ vs.  density. There is a large spread of densities within the dataset and the majority of the COFs possess a density of 0.45-0.55 $\mathrm{g \ cm^{-3}}$. Moreover, in accordance with previous findings \cite{thakur2024pushing}, we observe an increasing trend in $\kappa$ with increasing density. {The Pearson correlation coefficient between $\kappa$ and density is $r=0.594$, suggesting a moderate positive relationship.} However, increasing the density increases the ceiling of attainable $\kappa$, but does not guarantee a high value. 

\textbf{Pore size:} The pore size of any COF is represented by the largest pore diameter (LPD). \figurename\ref{fig:distribution}b shows that LPD in the dataset is spread over a large range between 5 \AA~and 80 \AA~and a concentration of structures in the 15-25 \AA~region. Moreover, there is an inverse relationship between the pore size and the thermal conductivity, as previously discovered by Freitas et al. \cite{freitas2017thermal}, where the range of achievable $\kappa$ decreases significantly with increasing pore sizes. {This is evidenced by the correlation coefficient $r=-0.508$, indicating a moderate negative correlation.} Here again, a small pore size does not guarantee a high thermal conductivity. On the other hand, the pore size can be taken as the sole design parameter to effectively reduce the thermal conductivity range (as mentioned in refs. \cite{thakur2024pushing,islamov2023high}). 

\textbf{Void fraction:} \figurename\ref{fig:distribution}c shows that the distribution of the void fractions of COFs across the dataset are within the 0.5–0.96 range, with the biggest share of COFs possessing a void fraction of around 0.85 and higher. Furthermore, a high void fraction of around 0.85 delivers both the largest values and range of attainable $\kappa$ within our dataset.  This finding is counterintuitive. Void fraction is typically inversely correlated with density (see Supplementary Information Section 1.3). Although low density is generally associated with a low $\kappa$ value, as shown in \figurename\ref{fig:distribution}a, the unexpected observation is that a high void fraction (which corresponds to low density) surprisingly results in a high $\kappa$ value, as also depicted in \figurename\ref{fig:distribution}c. {Beyond a favorable range of void fractions, no distinct trend emerges in the relationship with $\kappa$, as the correlation coefficient $r=-0.213$ suggests only a weak negative relationship.}

\textbf{Surface area:} Gravimetric surface area (GSA), i.e., surface area per unit mass varies highly across all the COF structures (see \figurename\ref{fig:distribution}d) with a larger cluster of points in the 7000-8000 $\mathrm{m^2 \ g^{-1}}$ range. We also observe that around intermediate GSA of 7,000 $\mathrm{m^2 \ g^{-1}}$, we have the widest spread and highest values of $\kappa$. {The correlation coefficient of $r=-0.147$ indicates a very weak negative relationship, suggesting that GSA is not a significant predictor.}

From the above initial analysis, we notice that the geometrical descriptors used here are insufficient to provide solid trends with respect to thermal conductivity. For example, with all the optimal descriptor values, i.e., a large density, a low pore size, a large void fraction, and an intermediate GSA, we are still not guaranteed to obtain a COF with a high $\kappa$. To elucidate this further, we choose four COFs -- denoted by (i)-(iv) -- and compare them in \figurename\ref{fig:distribution}a-d. 

COF~(i) has the highest thermal conductivity in the dataset with $\kappa =4.025 \ \mathrm{W m^{-1} K^{-1}}$. The structure has a relatively small pore size with 13.71 \AA~(see \figurename\ref{fig:distribution}a), a medium density with 0.928 $\mathrm{g \ cm^{-3}}$ (see \figurename\ref{fig:distribution}b), a relatively high void fraction of 0.844 (see \figurename\ref{fig:distribution}c), and an intermediate surface area of 7381 $\mathrm{m^2 g^{-1}}$. In contrast, the second COF~(ii) possesses a similar pore size of 9.88 \AA, a similar density of 0.938 $\mathrm{g \ cm^{-3}}$, a similar void fraction of 0.869, and a similar surface area of 7171 $\mathrm{m^2 g^{-1}}$, but exhibits thermal transfer capabilities almost seven times lower than that of (i) with $\kappa=0.603 \ \mathrm{W m^{-1} K^{-1}}$. 

The third COF (iii) has a pore size of 33.73 \AA~in the intermediate-to-high range, a low density at 0.408 $\mathrm{g \ cm^{-3}}$, a high void fraction at 0.932, and a slightly larger surface area of 8741 $\mathrm{m^2 g^{-1}}$. Following the trends described in literature \cite{thakur2024pushing,freitas2017thermal}, the geometrical descriptors are not optimal for a high $\kappa$. Nonetheless, COF (iii) has rather high $\kappa=1.96 \ \mathrm{W m^{-1} K^{-1}}$.  Lastly, COF (iv) has an exceptionally high pore size at 80.68 \AA~and a low density of 0.262 $\mathrm{g \ cm^{-3}}$ but  high void fraction of 0.958. Due to the extreme geometry, the COF (iv) has a very low thermal conductivity of $\kappa=0.289 \ \mathrm{W m^{-1} K^{-1}}$.

{Furthermore, we trained classical ensemble regression models using these descriptors to predict thermal conductivity (detailed results provided in Supplementary Information Section 2) and observed that the models yield poor prediction accuracy. 
In summary, the correlation analysis with the previously introduced descriptors and the presented examples illustrate the fact that the thermal conductivity structure-property relationships are complex and call for further examination.}

\subsection*{Predicting thermal conductivity using deep learning}

\begin{figure*}[t]
\centering
\includegraphics[width=\linewidth]{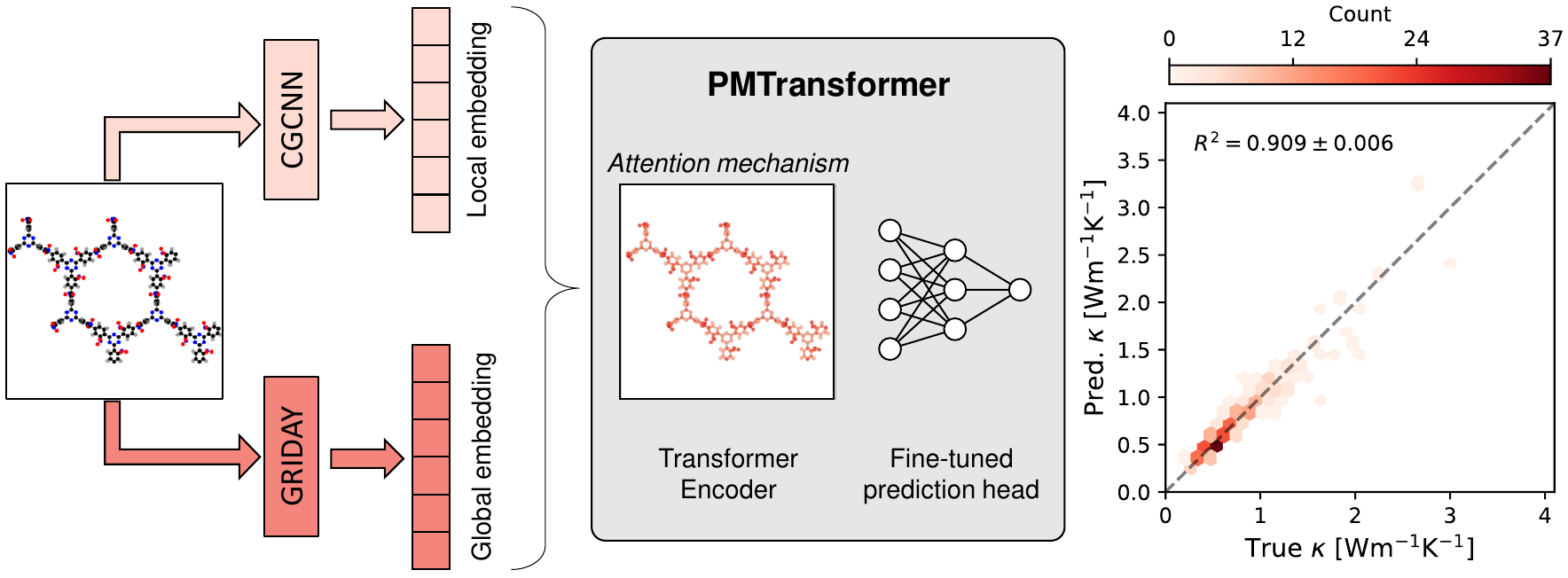}
\caption{Schematic of PMTransformer model. A sample COF structure is shown on the left. A crystal graph convolutional neural network computes local embeddings of the COF graph, while GRIDAY computes a three-dimensional energy grid of the structure, which becomes the global embeddings. These embeddings are combined and input into the PMTransformer (which includes a transformer encoder with attention mechanism and a prediction head) to predict $\kappa$. On the right, a parity plot compares $\kappa$ predictions with ground truth values. {We report the mean and standard deviation of the goodness-of-fit across five random seeds. The random seeds affect the initialization of the prediction head, while the rest of the model is initialized from the pre-trained PMTransformer weights. }{The dashed line represents the ideal line with zero intercept and unit slope.} }
\label{fig:ML_framework}
\end{figure*}

Our starting point is the deep learning-based Porous Material Transformer (PMTransformer) developed by Park et al. \cite{park2023enhancing} The PMTransformer is a multi-modal transformer \cite{vaswani2017attention} model (see \figurename\ref{fig:ML_framework} for schematic) designed for universal transfer learning for predicting the properties of porous materials, including COFs, MOFs, and zeolites. It receives two distinct sets of inputs: local and global features. Local features, which reflect the chemistry of building blocks and bonds, are derived from a crystal graph convolutional neural network \cite{xie2018crystal} (CGCNN). The CGCNN directly processes the crystal graph, where atoms and bonds are nodes and edges, respectively, and generates embeddings that capture essential chemical details. These embeddings serve as local features in the PMTransformer. Conversely, global features describe crystalline characteristics, including topological and geometric descriptors such as pore size and surface area. These features are derived from three-dimensional energy grids, which are generated by calculating the interaction energy between the material structure and a methane gas molecule at each grid point using the package GRIDAY \cite{githubGitHubSangwon91GRIDAY}. Similar to images in Vision Transformers \cite{alexey2020image}, the energy grids are divided into patches and flattened through linear projections. Eventually, both local and global embeddings are input into a transformer encoder, where the attention mechanism \cite{vaswani2017attention} helps the model focus on the most relevant parts of the molecular structure for predicting (in our case) the thermal conductivity.

To enable universal transfer learning, the transformer encoder in PMTransformer is pre-trained on an extremely large dataset of 1.9 million hypothetical porous materials to predict easily obtainable yet essential properties such as topology, void fraction, and building block prediction. This approach ensures the encoder captures critical information necessary for accurately predicting other, more complex properties in downstream tasks with much smaller datasets. For more details regarding the pre-training of PMTransfomer, refer to \cite{park2023enhancing}.

In this study, we leveraged the pre-trained transformer encoder of the PMTransformer model to perform transfer learning for the prediction of the thermal conductivity of COFs. 
We fine-tune the transformer model using the pre-trained weights as a starting point and jointly train a prediction head based on a multi-layer perceptron architecture to map the output of the transformer encoder to the thermal conductivity of COFs. We used the mean squared error (MSE) as a loss function for training. The dataset of 2,471 COFs is randomly split into two subsets: 90\% of the data is used for training (out of which 10\% for validation), and the remaining data is used for testing. {We repeat the training across five different random seeds, where each seed initializes the prediction head randomly while keeping the transformer initialized with the same pre-trained weights.} Additional details of the training protocol {and an ablation study} are provided in Supplementary Information Section 3. 

{We demonstrate the model's prediction accuracy on the test dataset in \figurename\ref{fig:ML_framework}, where it achieves a goodness-of-fit $R^2$ of $0.909\pm0.006$ and a mean absolute error (MAE) of $0.075 \ \mathrm{W m^{-1} K^{-1}}$ for predicting $\kappa$. The model's strong performance suggest that there may be gaps in the feature sets typically used to describe COFs. It also highlights the possibility that additional key descriptors, beyond those commonly used, could play an important role in predicting thermal conductivity, thus warranting further exploration of structure-property relationships.}

\subsection*{Discovering novel thermal transfer mechanisms for {two-dimensional COFs} via self-attention} 

We leverage the multi-head self-attention mechanism \cite{vaswani2017attention} of the transformer architecture to interpret the predictions made by the deep learning model. The attention mechanism in transformers is designed to identify and assign weights, known as attention scores, to the significance of different parts of the input data/encodings relative to all the other parts. When applied to molecular structures, such as COFs, the attention mechanism enables the model to dynamically focus attention on key  molecular substructures in relation to all the other substructures present based on their relevance to the property being predicted, in this case, thermal conductivity. In multi-head attention, multiple attention heads are needed to capture different aspects or relationships within the data simultaneously. Specifically, in the context of the PMTransformer model applied to a COF to predict its thermal conductivity, we calculate the attention for each atom in the COF by averaging the attention scores from all attention heads and computing the joint attention via multiplicative aggregation across all layers of the transformer encoder. 

\begin{figure*}[h!]
\centering
\includegraphics[width=\linewidth]{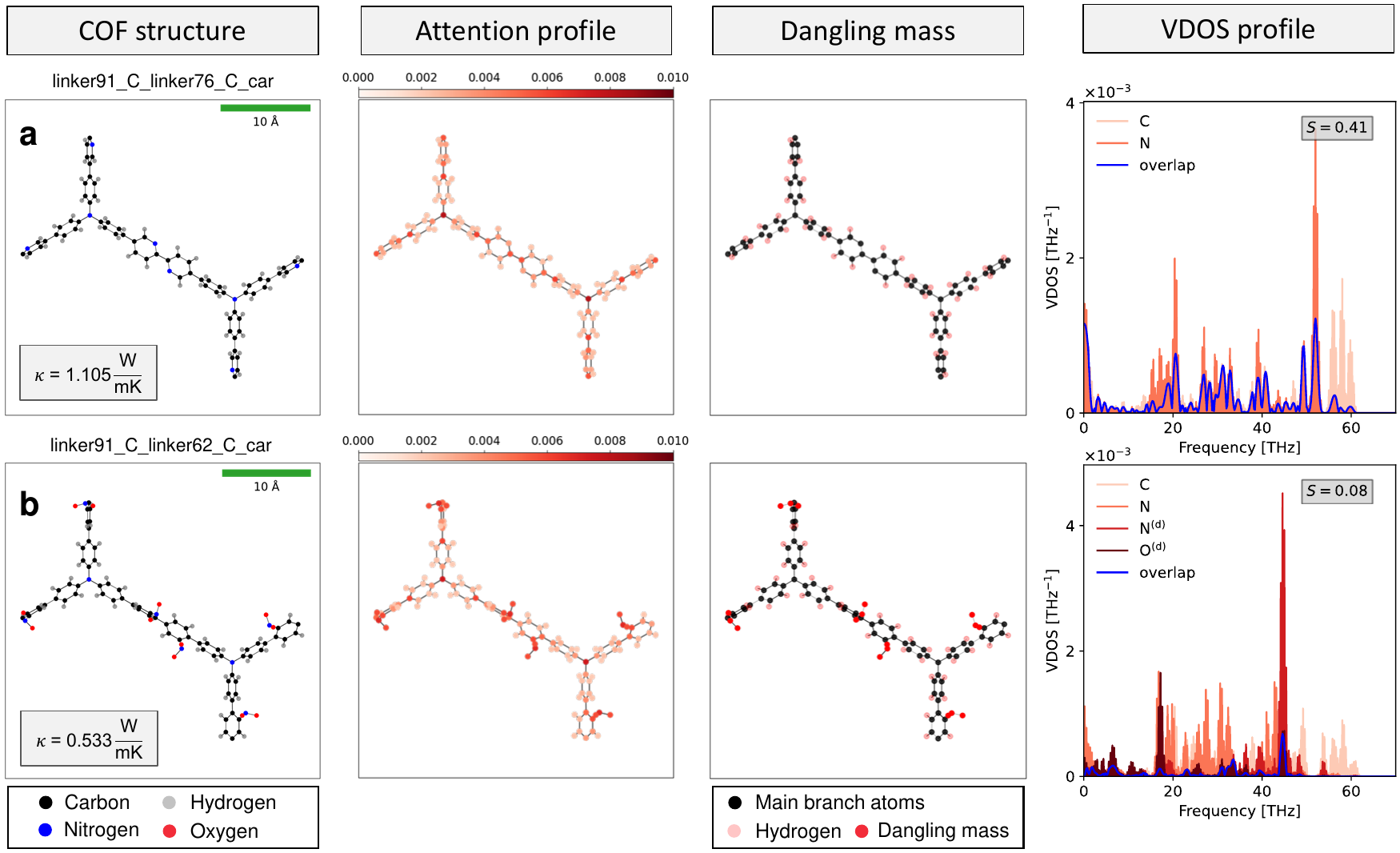}
\caption{An example pair of COFs with same topologies, similar geometric descriptors, but contrasting  thermal conductivities. The first column illustrates the COF structures. The second column shows the atom-level attention score profile computed by the attention mechanism. The third column shows the same COF structure distinguishing atoms on the main branch and dangling atoms (with separate distinction for hydrogen atoms). The fourth column shows the VDOS profiles of various groups of atoms within the corresponding COF structure with the overlap metric $S$. The legend indicates the VDOS profile for main branch atoms $(.)$ and dangling atoms $(.)^{\text{(d)}}$. {The reported thermal conductivities are obtained from NEMD simulations, rather than predicted by the PMTransformer model.}}
\label{fig:vdos_dangling1}
\end{figure*}

\begin{figure*}[h!]
\centering
\includegraphics[width=\linewidth]{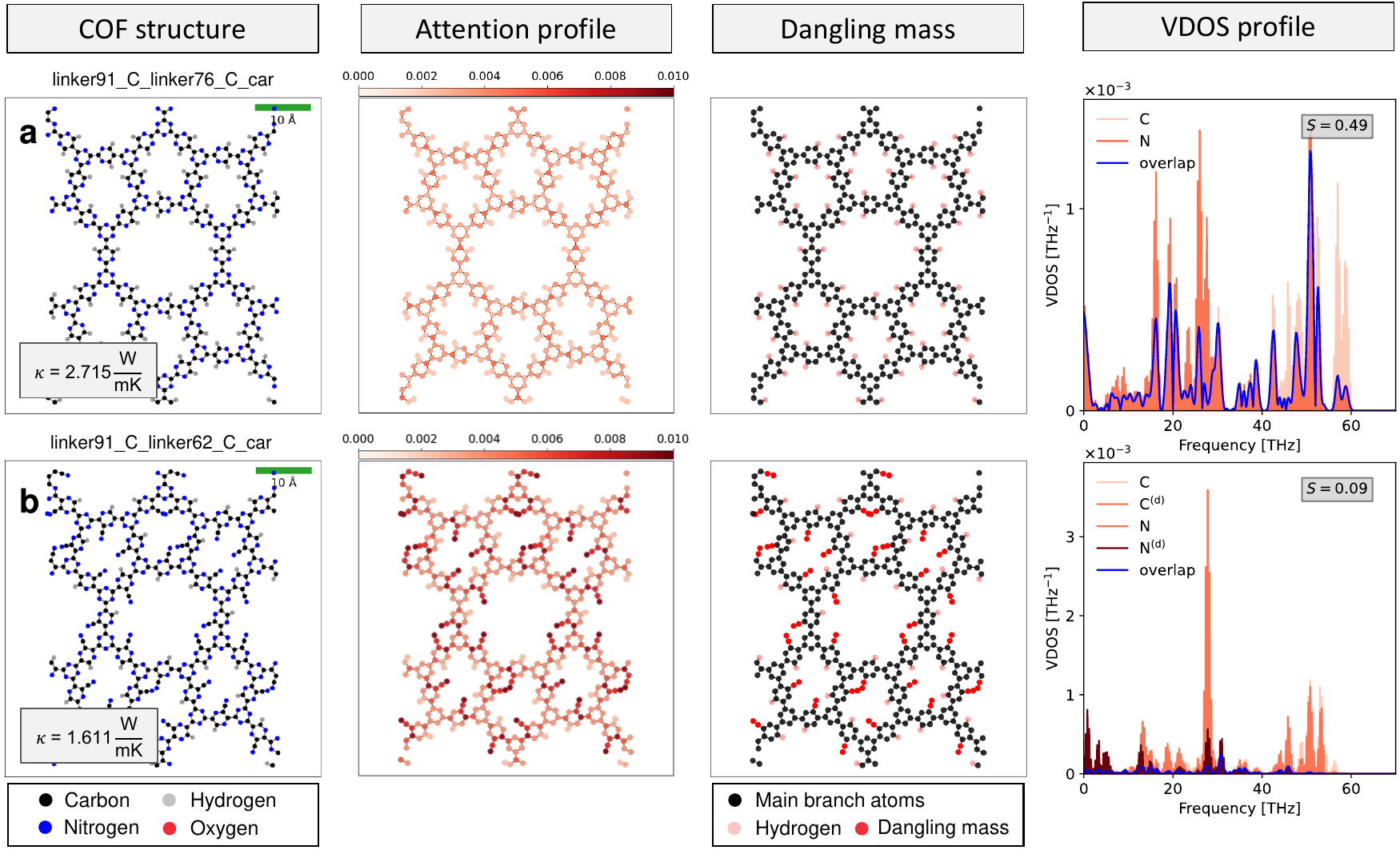}
\caption{An example pair of COFs with same topologies, similar geometric descriptors, but contrasting  thermal conductivities. The first column illustrates the COF structures. The second column shows the atom-level attention score profile computed by the attention mechanism. The third column shows the same COF structure distinguishing atoms on the main branch and dangling atoms (with separate distinction for hydrogen atoms). The fourth column shows the VDOS profiles of various groups of atoms within the corresponding COF structure with the overlap metric $S$. The legend indicates the VDOS profile for main branch atoms $(.)$ and dangling atoms $(.)^{\text{(d)}}$. {The reported thermal conductivities are obtained from NEMD simulations, rather than predicted by the PMTransformer model.}}
\label{fig:vdos_dangling2}
\end{figure*}

Additionally, we identify the \textit{main branch} of a COF as the shortest continuous path connecting the  boundary points necessary for periodicity. We then classify all the atoms extending and excluded from the main branch as \textit{dangling mass}. More details on how we classify each atom as either part of the main branch or part of a dangling side branch are presented in Supplementary Information Section 4.  We observe distinct patterns in the attention assigned to various atomic sites within a COF's graph representation, which correspond to the location of the dangling masses. Two representative examples are shown in \figurename\ref{fig:vdos_dangling1} and \figurename\ref{fig:vdos_dangling2}. 

Consider the first example (see \figurename\ref{fig:vdos_dangling1}a-b). The top row illustrates a COF structure (\figurename\ref{fig:vdos_dangling1}a) from the test dataset with a thermal conductivity of $1.105 \ \mathrm{W m^{-1} K^{-1}}$ and no dangling masses except hydrogen atoms. Additionally, the atom-wise attention scores are uniformly distributed and low for all the carbon atoms with some elevated attention to the nitrogen atoms. In the second row, we illustrate a COF structure (\figurename\ref{fig:vdos_dangling1}b) with the same topology as the previous one, but with a much lower thermal conductivity of  $0.533 \ \mathrm{W m^{-1} K^{-1}}$. Its attention profile reveals that certain atom groups are being paid special attention to by the transformer, i.e., specifically the $-\text{NO}_2$ functional group on the benzene ring that is dangling from the main branch of the COF. Notably, the same groups of atoms that exhibit higher attention scores are also classified as dangling masses (see \figurename\ref{fig:vdos_dangling1}b attention profile \& dangling mass). 

In the second example (\figurename\ref{fig:vdos_dangling2}a-b), the COF structure on the top does not contain any dangling mass except hydrogen atoms and has a thermal conductivity of $2.715 \ \mathrm{W m^{-1} K^{-1}}$ with a highly uniform attention profile (see \figurename\ref{fig:vdos_dangling2}a). Analogously, we pick a second COF structure with the same topology and comparable geometrical descriptors. Once more, we observe that the second structure contains a significant amount of dangling mass (i.e., $-\text{CN}$ branches extending from the rings) with corresponding elevated attention scores and has a lower thermal conductivity of $1.611 \ \mathrm{W m^{-1} K^{-1}}$.

The same trend is observed for numerous examples, with additional ones presented in the Supplementary Information Section 5. We {hypothesize} that this increased attention is indicative of the deep learning model's understanding that these dangling masses are significant in predicting thermal conductivity. Furthermore, it suggests that the presence of dangling masses disrupts the heat transfer pathways and thereby reduces the thermal conductivity through the material. A similar effect of dangling mass lowering the thermal conductivity has been reported previously in singular polymer chains \cite{luo2018decreased} and amorphous polymers \cite{feng2020size}. {This effect, while known in disordered systems, has not been examined in crystalline COFs.}
{Here, we demonstrate that it remains relevant even in ordered, porous frameworks, inviting future investigation with high-fidelity modeling methods (such as density functional theory) to examine the underlying mechanism in more detail.}

To further validate the structure-property relationship interpreted from the deep learning model, we examine the impact of dangling atoms on the vibrational density of states (VDOS) for the aforementioned contrasting COF examples. VDOS, which characterizes the distribution of vibrational modes in a system as a function of frequency, is known to impact thermal transfer properties. Overlaps in VDOS profiles between different atoms have been shown to affect these properties in both COFs \cite{feng2020thermal} and MOFs \cite{ying2021effect}.

Through MD simulations, the VDOS is calculated using the Fourier transform of the normalized velocity autocorrelation function of specific groups of atoms (see Supplementary Information Section 6 for details). \figurename\ref{fig:vdos_dangling1}  and \figurename\ref{fig:vdos_dangling2}(third column) show the VDOS profile of carbon, nitrogen, and oxygen atoms in the representative COF examples. For any COF, we define a VDOS overlap metric $S$  as the ratio of the area under the curve (AUC) of the minimum VDOS across all atom types (at each frequency) and the AUC of the maximum VDOS across all atom types (at each frequency), i.e.,
\begin{align}
\begin{split}
    S = \frac{\int_{\omega} \min\left\{\ \mathrm{f^{V}_{C}(\omega),\ f^{V}_{N}(\omega),\ f^{V}_{O}(\omega),\ f^{V}_{B}(\omega),\ f^{V,d}_{C}(\omega),\ f^{V,d}_{N}(\omega),\ f^{V,d}_{O}(\omega),\ f^{V,d}_{B}(\omega) \ }\right\} \dd \omega}{\int_{\omega} \max\left\{\ \mathrm{f^{V}_{C}(\omega),\ f^{V}_{N}(\omega),\ f^{V}_{O}(\omega),\ f^{V}_{B}(\omega),\ f^{V,d}_{C}(\omega),\ f^{V,d}_{N}(\omega),\ f^{V,d}_{O}(\omega),\ f^{V,d}_{B}(\omega)}\right\} \dd \omega}.
    \end{split}
\end{align}

Here, $\omega$ denotes the frequency; $\mathrm{f_C^{V}}(\omega)$, $\mathrm{f_N^{V}}(\omega)$, $\mathrm{f_O^{V}}(\omega)$, and $\mathrm{f_B^{V}}(\omega)$ denote the  VDOS at frequency $\omega$ for the carbon, nitrogen, oxygen and boron atoms, respectively. Additionally, we differentiate between atoms in the main branch and dangling atoms, where the VDOS profiles for dangling atoms are denoted with a $(.)^{\text{d}}$ superscript. A large overlap in VDOS profiles of different atom types would result in $S\approx 1$, whereas a small overlap would yield $S\approx 0$. A high overlap indicates that phonon waves (i.e., quantized modes of vibrations responsible for thermal energy transfer within the crystal lattice) can propagate freely throughout the structure. This would lead to minimized phonon scattering and facilitate efficient heat transfer across the material \cite{wang2023molecular}.

In the COF structures, which lack or have minimal dangling atoms, the VDOS profiles of the atoms are broad and overlap significantly, e.g., $S\approx 0.41$ and $0.49$ for examples in \figurename\ref{fig:vdos_dangling1}a  and \figurename\ref{fig:vdos_dangling2}a, respectively. This results in an even distribution of vibrational modes across a wide frequency spectrum for all atoms.  In contrast, the COFs with substantial dangling atoms exhibit a VDOS profile with minimal overlap, e.g., $S\approx 0.08$ and $0.09$ for examples in \figurename\ref{fig:vdos_dangling1}b  and \figurename\ref{fig:vdos_dangling2}b, respectively. {Notably, dangling atoms significantly reduce the VDOS overlap in both low-frequency (0 to 20 THz) and high-frequency (40 to 60 THz) modes.} The vibrational modes of individual atoms in these COFs are confined to much narrower frequency bands, which leads to a less harmonized VDOS profile. The lack of overlap suggests that the vibrations are highly localized around the dangling atoms, which causes the phonon waves to be scattered and disrupting their ability to transfer heat efficiently through the material \cite{ying2020impacts}.
We interpret the mismatch in the VDOS profiles as an energy barrier for phonon transport, where phonons encounter resistance, preventing smooth energy transfer across atoms. This suggests that the absence of dangling mass is crucial for enhancing thermal conductivity in COFs. Dangling atoms create these mismatches in VDOS, which in turn hinders phonon transport and reduces thermal conductivity. {A similar trend is observed in additional pairs of COFs (\figurename S9, Supplementary Information) and the correlation between $\kappa$ and $S$ for a selected representative set of eight COFs is presented in \figurename S6c (Supplementary Information).}

To further confirm the effect of dangling atoms on phonon dispersion in COFs, we calculate the phonon spectral energy density (pSED) of the example COFs and their dangling counterparts, as presented in Section 5 in Supplementary Information. The pSED of COFs containing more dangling atoms exhibits higher magnitudes and broadening bands, indicating stronger anharmonicity and vibrational scattering in the vibrational modes, consequently resulting in reduced thermal conductivity.

To quantify the role of dangling masses in the structure-property relationship for thermal conductivity of COFs, we introduce the dangling mass ratio (DMR). The DMR $\in[0,1]$ quantifies the ratio between the mass of dangling atoms and the total mass of the COF, i.e.,
\begin{align}
\text{DMR} = \frac{\sum_{i\in \calD}m_{{i}}} {\sum_{i\in \calD}m_{{i}} + \sum_{i\in \calM}m_{{i}}}, 
\end{align}
where $m_i$ is the mass of the $i^\text{th}$ atom belonging to the set of either dangling atoms or of non-dangling main branch atoms -- denoted by $\calD$ and $\calM$, respectively. Higher DMR values indicate a greater proportion of dangling masses relative to the COF's total mass. {In Supplementary Information Section~1.3, we qualitatively show the relationship between DMR, the VDOS overlap $S$, and $\kappa$.}

\begin{figure*}[t]
\centering
\includegraphics[width=0.7\linewidth]{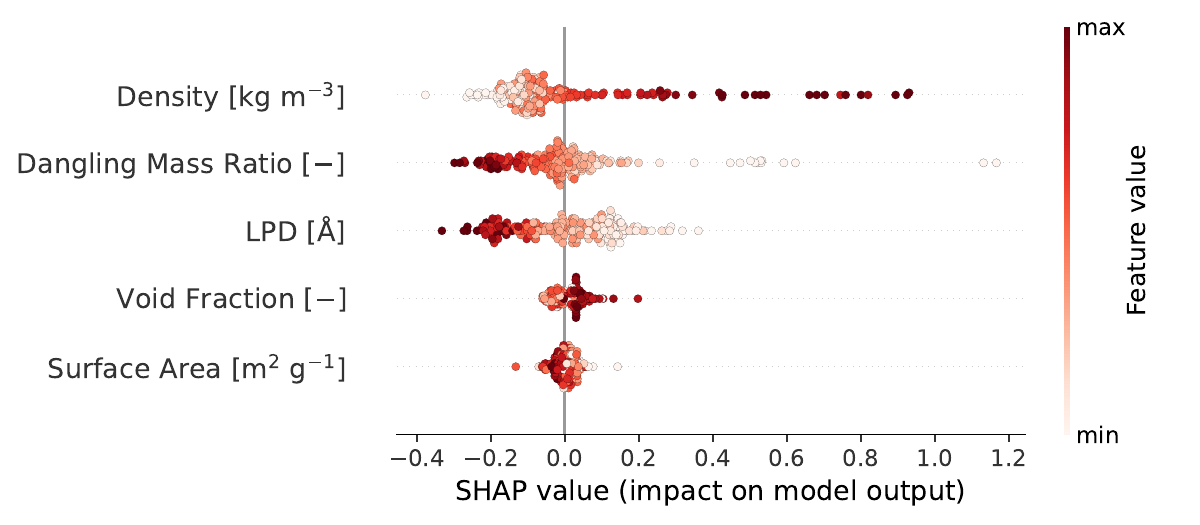}
\caption{{Distribution of the effect of each feature on the model's predictions. Each point represents a SHAP value for a feature across all samples, with the color intensity indicating feature values (ranging between minimum and maximum values). The position along the x-axis represents the magnitude of the feature's impact.}}
\label{fig:shap_analysis}
\end{figure*}

We then introduce DMR as an additional feature, alongside pore size, density, void fraction, and surface area, and train classical ensemble regression models{, such as the Random Forest \cite{breiman2001random}, Gradient Boosting \cite{friedman2001greedy}, XGBoost \cite{chen2016xgboost}, and AdaBoost \cite{freund1995desicion} and outline their performances in \tablename~\ref{tab:regression-performance}.} {These simpler models are employed not to compete with the PMTransformer in predictive accuracy, but to provide directly interpretable feature importance scores and understanding the relevance of DMR. In contrast, extracting meaningful and quantitative feature importance from the PMTransformer is non-trivial due to its complex model architecture and the absence of an explicit correspondence between input features and learned representations.} To assess the importance of individual features, including DMR, we analyzed their Gini importance {and conducted a permutation feature importance analysis \cite{scikit-learn}}. These values quantify the contribution of each feature to the overall predictive power of the model. The Gini importance measures how much each feature reduces the Gini impurity or randomness when making predictions in the ensemble model \cite{breiman2001random}. {On the other hand, permutation feature importance measures the impact of a feature by assessing how much the model's performance decreases when the values of that feature are randomly shuffled. A greater drop in performance indicates that the feature is more important in predicting the target variable \cite{breiman2001random}.}
{We note that for the best performing regressor, the Random Forest, DMR emerges as the second most influential predictor with 27.8\% Gini importance, closely following behind density (see \tablename~\ref{tab:importance}). Similarly, the permutation feature importance analysis also highlights DMR as a key feature, further supporting its significant role in predicting thermal conductivity, with its importance consistently ranking just below density.} {Finally, we conducted SHAP analysis \cite{lundberg2017shap} on the the Random Forest to better understand the impact of each feature on the individual test data points. As shown in \figurename\ref{fig:shap_analysis}, the analysis reveals that higher DMR values are associated with lower predicted $\kappa$, confirming the trend we observed. The consistency of these findings across different methods---Gini, permutation feature importance, and SHAP---reinforces the importance of DMR as a key factor in predicting thermal conductivity.} {We note that the computed importance scores are inherently linked to the design space and dataset introduced by Mercado et al. \cite{mercado2018silico}. Specifically, if the dataset did not contain structures with dangling masses, the DMR feature would naturally show little to no importance. Thus, the relevance of DMR as a descriptor depends on its variability within the dataset under consideration.}

\begin{table*}[t]
\centering
\begin{minipage}{0.48\linewidth}
\centering
{
\begin{tabular}{lc}
\toprule
\textbf{Regression Model} & \textbf{$R^2$} \\ 
\midrule
Random Forest              & $\mathbf{0.728\pm0.069}$ \\ 
Gradient Boosting          & $0.688\pm0.070$         \\ 
XGBoost                    & $0.691\pm0.035$         \\ 
AdaBoost                   & $0.416\pm0.125$         \\ 
\bottomrule
\end{tabular}
}
\caption{{Performances of standard ensemble regression models at predicting $\kappa$ by including DMR as an input feature. The table reports $R^2$ scores obtained through 10-fold cross-validation.}}
\label{tab:regression-performance}
\end{minipage}
\hfill
\begin{minipage}{0.48\linewidth}
\centering
\caption*{{Feature importances for Random Forest}}
{
\begin{tabular}{lcc}
\toprule
\textbf{Feature}        & \textbf{Gini} & \textbf{PFI} \\ 
\midrule
Density                 & 0.397         & 0.670        \\ 
\textbf{DMR}                     & \textbf{0.278}         & \textbf{0.511 }       \\ 
LPD                     & 0.185         & 0.455        \\ 
Void Fraction           & 0.070         & 0.093        \\ 
GSA            & 0.069         & 0.022        \\ 
\bottomrule
\end{tabular}
}
\caption{{Feature importances for the best-performing regression model, i.e., Random Forest. The table reports both Gini and Permutation Feature Importance (PFI) values for all features.}}
\label{tab:importance}
\end{minipage}
\end{table*}

\subsection*{Generalizability and high-throughput screening for discovering {two-dimensional monolayer} COFs with tailored properties}
\begin{figure*}[t]
\centering
\includegraphics[width=\linewidth]{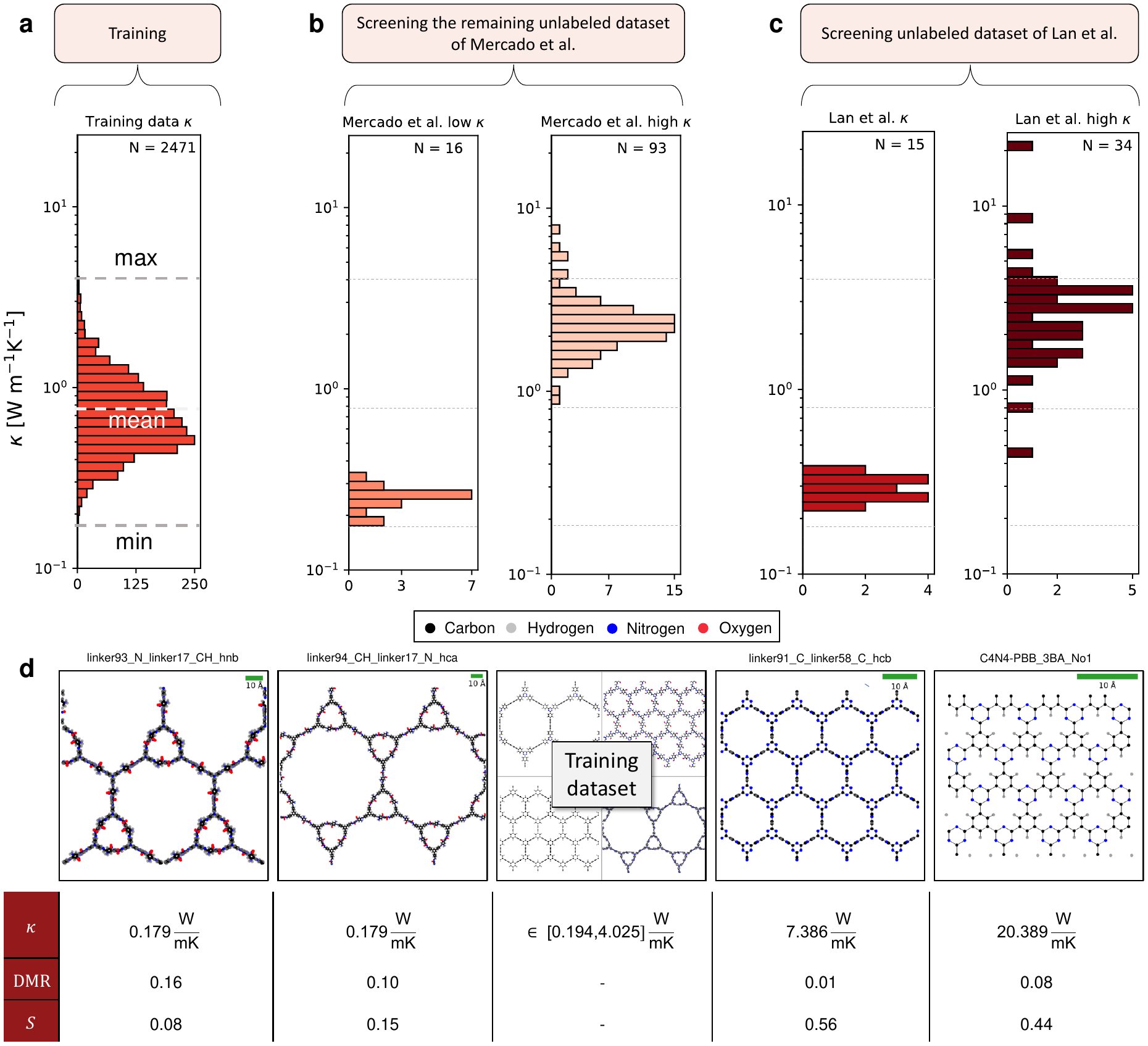}
\caption{High-throughput screening. \textbf{(a)} Distribution of $\kappa$ in the training dataset. Dashed lines indicated minimum, mean, and maximum $\kappa$ across the training dataset. \textbf{(b,c)} Distribution of $\kappa$ of selected COF candidates identified in high-throughput screening for low and high $\kappa$ in the unlabeled datasets of {(b)} Mercado et al. \cite{mercado2018silico} and {(c)} Lan et al. \cite{lan2018materials} While the former dataset shares the same design space as the training dataset, the latter has a different design space. All thermal conductivities shown here are computed using MD. \textbf{(d)} Representative COFs identified through high-throughput screening with $\kappa$ lower and  higher than the minimum and maximum, respectively, across the training dataset. Also indicated are their corresponding DMR and $S$ values.}
\label{fig:screening}
\end{figure*}

To evaluate the generalization capabilities of the deep learning model for high-throughput screening, we screen previously unseen 6,170 two-dimensional COFs in the unlabelled dataset by Mercado et al. \cite{mercado2018silico} from which our training and test sets were originally subsampled. Moreover, we screen $35,638$ two-dimensional COFs in the unlabelled dataset by Lan et al. \cite{lan2018materials} This dataset covers a vastly different design space with a variety of linkers and knots previously unseen by our trained model.

The screening is performed with the objective of facilitating the identification COFs with extremely high or low thermal conductivity without resorting to computationally expensive simulations. We observe an average screening rate of 0.07 seconds per COF structure (see Supplementary Information Section 7 for details on computational runtimes), thereby enabling a speedup of almost seven orders of magnitude compared to MD-based screening. 

Once potential candidates with extreme thermal conductivities are identified using the PMTransformer, we then calculate their thermal conductivities through MD simulations for these selected few, rather than for the entire datasets. {This approach circumvents reliance on the PMTransformer's absolute predictions by validating promising candidates with high-fidelity MD calculations.} The results are summarized in \figurename\ref{fig:screening}. Our screening process on the dataset by Mercado et al. \cite{mercado2018silico} revealed COFs exhibiting thermal conductivities up to twice the maximum value observed in the training set, but stemming from the same design space. By screening the dataset by Lan et al. \cite{lan2018materials}, we discover COFs with thermal conductivities five times as high as the ones in our training set. Additionally, we discovered several COFs with thermal conductivities lower than any values previously recorded in our dataset. A visual inspection reveals that these COFs feature a combination of large pore sizes, low density, and a substantial presence of dangling masses. For all the identified COFs, we observe that thermal conductivity shows a strong trend with both dangling mass ratio (DMR) and VDOS overlap ratio ($S$), which is in agreement with the previous observations.

\section*{Discussion}

In this study, we compiled an extensive dataset of thermal conductivities for 2,471 two-dimensional COFs, computed using NEMD. Despite observing some structure-property trends  with conventional descriptors such as density, pore size, void fraction, and surface area, no single descriptor or combination thereof consistently predicted thermal conductivity with high accuracy, highlighting the complexity of COF structure-property relationships. To enhance prediction accuracy, we trained a transformer-based deep learning model incorporating a multi-head attention mechanism. This model significantly outperformed traditional ensemble regression models, achieving an $\mathrm{R^2}$ of $0.909\pm0.006$.

{Further analysis using the transformer's attention mechanism revealed that COFs with higher amounts of dangling atoms exhibited lower thermal conductivities due to disrupted heat transfer pathways, identifying a novel and significant predictor of thermal conductivity. The random forest regressor, identified as the best-performing regression ensemble model, {was used primarily as a simple tool to assess feature importance rather than to compare predictive performance with the PMTransformer.} It was analyzed using Gini importance, permutation feature importance, and SHAP values. These analyses consistently highlighted the dangling mass ratio as the second most important predictor of thermal conductivity. The vibrational density of states analysis further supported these findings, showing that dangling masses introduce mismatched vibrational modes that hinder thermal transfer.} {While our current analysis focuses on two-dimensional monolayer COFs, additional mechanisms may emerge in three-dimensional architectures and warrant separate investigation.}

We further leveraged the deep learning model to efficiently screen thousands of COFs, both within and beyond the design space we analyzed, to identify candidates with extreme thermal conductivities. This approach provides a rapid and reliable method as well as valuable insights for designing COFs with tailored thermal characteristics. To support further research, we release the dataset of COF thermal conductivities and encourage the research community to utilize and expand upon it. Finally, we emphasize the need to bridge high interpretability with high accuracy in machine learning models to navigate the complex structure-property space of COFs and other nanoporous materials, thereby enabling the design of optimal materials for applications such as gas separation and thermal management.

\section*{Methods}

The molecular dynamics setup, the distribution of in-plane thermal conductivities, and feature correlations (including DMR) (Section 1); implementation details of the ensemble regression models and regression performances excluding DMR (Section 2); the deep learning setup (Section 3); the dangling mass computation (Section 4); additional comparisons of COFs with and without dangling mass, including phonon dispersion maps for all examples (Section 5); details on the VDOS and pSED computation (Section 6); and estimated runtimes (Section 7) are summarized in the Supplementary Information.

\section*{Author contributions}
\textbf{P.T and Y.Z.:} conceptualization, data curation, formal analysis, investigation, writing -- original draft. \textbf{P.T., Y.Z., Y.G.:} methodology, software, validation, visualization. \textbf{A.V. and S.K.:} conceptualization, supervision, funding acquisition, writing -- review \& editing

\section*{Competing interests}
The authors declare no competing interests.

\section*{Data and code availability}
The code and data generated during this study are available in the following repository: \href{https://doi.org/10.4121/5866cc9a-78bf-4a0c-9280-ae526da86ac9}{doi.org/10.4121/5866cc9a-78bf-4a0c-9280-ae526da86ac9}. The code and data are also available at: \href{https://github.com/mmc-group/deep-learning-thermal-conductivity-of-COFs}{github.com/mmc-group/deep-learning-thermal-conductivity-of-COFs}.
The data includes unit-cell CIF files obtained from the dataset of Mercado et al.\cite{mercado2018silico} available at \href{https://doi.org/10.24435/materialscloud:2018.0003/v2}{doi.org/10.24435/materialscloud:2018.0003/v2}.
We also used the dataset provided by Lan et al.\cite{lan2018materials}, which is available at~\href{https://figshare.com/s/c7e3b7610a71b9d64210}{figshare.com/s/c7e3b7610a71b9d64210} with DOI: 10.1038/s41467-018-07720-x for the associated paper.
The codes are built upon the open-source MOFTransformer/PMTransformer repository by Park et al.\cite{park2023enhancing}, which can be accessed at \href{https://github.com/hspark1212/MOFTransformer}{github.com/hspark1212/MOFTransformer} with DOI: 10.1021/acsami.3c10323 for the associated paper.

\section*{Acknowledgements}
P.T., Y.G., and S.K. acknowledge that this material is based upon work supported by the Air Force Office of Scientific Research under award number FA8655-23-1-7020. P.T., Y.G., and S.K. acknowledge that this work was supported in part by Oracle Cloud credits and related resources provided by Oracle for Research. A.V. and Y.Z. acknowledge  Microsoft Climate Research Initiative and University of Washington for funding. The funders played no role in study design, data collection, analysis and interpretation of data, or the writing of this manuscript.

We thank Prof. Marcel Sluiter of TU Delft for his helpful discussions and insights during the course of this research. Additionally, we thank Eric Vo of University of Washington for his help in preparing simulations of COF structures.

\balance

\bibliography{rsc} %

\providecommand*{\mcitethebibliography}{\thebibliography}
\csname @ifundefined\endcsname{endmcitethebibliography}
{\let\endmcitethebibliography\endthebibliography}{}
\begin{mcitethebibliography}{65}
\providecommand*{\natexlab}[1]{#1}
\providecommand*{\mciteSetBstSublistMode}[1]{}
\providecommand*{\mciteSetBstMaxWidthForm}[2]{}
\providecommand*{\mciteBstWouldAddEndPuncttrue}
  {\def\EndOfBibitem{\unskip.}}
\providecommand*{\mciteBstWouldAddEndPunctfalse}
  {\let\EndOfBibitem\relax}
\providecommand*{\mciteSetBstMidEndSepPunct}[3]{}
\providecommand*{\mciteSetBstSublistLabelBeginEnd}[3]{}
\providecommand*{\EndOfBibitem}{}
\mciteSetBstSublistMode{f}
\mciteSetBstMaxWidthForm{subitem}
{(\emph{\alph{mcitesubitemcount}})}
\mciteSetBstSublistLabelBeginEnd{\mcitemaxwidthsubitemform\space}
{\relax}{\relax}

\bibitem[Wang \emph{et~al.}(2020)Wang, Wang, Wang, Tang, Zeng, Xu, Chen, Xiong, Zhou, Li, Huang, Zhu, Wang, and Tang]{wang2020covalent}
H.~Wang, H.~Wang, Z.~Wang, L.~Tang, G.~Zeng, P.~Xu, M.~Chen, T.~Xiong, C.~Zhou, X.~Li, D.~Huang, Y.~Zhu, Z.~Wang and J.~Tang, \emph{Chemical Society Reviews}, 2020, \textbf{49}, 4135–4165\relax
\mciteBstWouldAddEndPuncttrue
\mciteSetBstMidEndSepPunct{\mcitedefaultmidpunct}
{\mcitedefaultendpunct}{\mcitedefaultseppunct}\relax
\EndOfBibitem
\bibitem[Zhao \emph{et~al.}(2021)Zhao, Pachfule, and Thomas]{zhao2021covalent}
X.~Zhao, P.~Pachfule and A.~Thomas, \emph{Chemical Society Reviews}, 2021, \textbf{50}, 6871–6913\relax
\mciteBstWouldAddEndPuncttrue
\mciteSetBstMidEndSepPunct{\mcitedefaultmidpunct}
{\mcitedefaultendpunct}{\mcitedefaultseppunct}\relax
\EndOfBibitem
\bibitem[Liu \emph{et~al.}(2021)Liu, Wang, Wan, and Yu]{liu2021mof}
C.~Liu, J.~Wang, J.~Wan and C.~Yu, \emph{Coordination Chemistry Reviews}, 2021, \textbf{432}, 213743\relax
\mciteBstWouldAddEndPuncttrue
\mciteSetBstMidEndSepPunct{\mcitedefaultmidpunct}
{\mcitedefaultendpunct}{\mcitedefaultseppunct}\relax
\EndOfBibitem
\bibitem[Stock and Biswas(2011)]{stock2012synthesis}
N.~Stock and S.~Biswas, \emph{Chemical Reviews}, 2011, \textbf{112}, 933–969\relax
\mciteBstWouldAddEndPuncttrue
\mciteSetBstMidEndSepPunct{\mcitedefaultmidpunct}
{\mcitedefaultendpunct}{\mcitedefaultseppunct}\relax
\EndOfBibitem
\bibitem[Pérez-Botella \emph{et~al.}(2022)Pérez-Botella, Valencia, and Rey]{perez2022zeolites}
E.~Pérez-Botella, S.~Valencia and F.~Rey, \emph{Chemical Reviews}, 2022, \textbf{122}, 17647–17695\relax
\mciteBstWouldAddEndPuncttrue
\mciteSetBstMidEndSepPunct{\mcitedefaultmidpunct}
{\mcitedefaultendpunct}{\mcitedefaultseppunct}\relax
\EndOfBibitem
\bibitem[Abuzeid \emph{et~al.}(2021)Abuzeid, EL-Mahdy, and Kuo]{abuzeid2021covalent}
H.~R. Abuzeid, A.~F. EL-Mahdy and S.-W. Kuo, \emph{Giant}, 2021, \textbf{6}, 100054\relax
\mciteBstWouldAddEndPuncttrue
\mciteSetBstMidEndSepPunct{\mcitedefaultmidpunct}
{\mcitedefaultendpunct}{\mcitedefaultseppunct}\relax
\EndOfBibitem
\bibitem[Dogru \emph{et~al.}(2013)Dogru, Handloser, Auras, Kunz, Medina, Hartschuh, Knochel, and Bein]{dogru2013photoconductive}
M.~Dogru, M.~Handloser, F.~Auras, T.~Kunz, D.~Medina, A.~Hartschuh, P.~Knochel and T.~Bein, \emph{Angewandte Chemie International Edition}, 2013, \textbf{52}, 2920–2924\relax
\mciteBstWouldAddEndPuncttrue
\mciteSetBstMidEndSepPunct{\mcitedefaultmidpunct}
{\mcitedefaultendpunct}{\mcitedefaultseppunct}\relax
\EndOfBibitem
\bibitem[Stegbauer \emph{et~al.}(2018)Stegbauer, Zech, Savasci, Banerjee, Podjaski, Schwinghammer, Ochsenfeld, and Lotsch]{stegbauer2018tailor}
L.~Stegbauer, S.~Zech, G.~Savasci, T.~Banerjee, F.~Podjaski, K.~Schwinghammer, C.~Ochsenfeld and B.~V. Lotsch, \emph{Advanced Energy Materials}, 2018, \textbf{8}, 1703278\relax
\mciteBstWouldAddEndPuncttrue
\mciteSetBstMidEndSepPunct{\mcitedefaultmidpunct}
{\mcitedefaultendpunct}{\mcitedefaultseppunct}\relax
\EndOfBibitem
\bibitem[Ali \emph{et~al.}(2024)Ali, Huo, Zhang, and Wang]{ali2024thiazole}
Z.~Ali, T.~Huo, Y.~Zhang and G.~Wang, \emph{Microchemical Journal}, 2024, \textbf{200}, 110340\relax
\mciteBstWouldAddEndPuncttrue
\mciteSetBstMidEndSepPunct{\mcitedefaultmidpunct}
{\mcitedefaultendpunct}{\mcitedefaultseppunct}\relax
\EndOfBibitem
\bibitem[Li \emph{et~al.}(2020)Li, Chen, Han, Feng, Zhang, and Zhang]{li2020fabrication}
Y.~Li, M.~Chen, Y.~Han, Y.~Feng, Z.~Zhang and B.~Zhang, \emph{Chemistry of Materials}, 2020, \textbf{32}, 2532–2540\relax
\mciteBstWouldAddEndPuncttrue
\mciteSetBstMidEndSepPunct{\mcitedefaultmidpunct}
{\mcitedefaultendpunct}{\mcitedefaultseppunct}\relax
\EndOfBibitem
\bibitem[Yusran \emph{et~al.}(2020)Yusran, Li, Guan, Fang, and Qiu]{yusran2020covalent}
Y.~Yusran, H.~Li, X.~Guan, Q.~Fang and S.~Qiu, \emph{EnergyChem}, 2020, \textbf{2}, 100035\relax
\mciteBstWouldAddEndPuncttrue
\mciteSetBstMidEndSepPunct{\mcitedefaultmidpunct}
{\mcitedefaultendpunct}{\mcitedefaultseppunct}\relax
\EndOfBibitem
\bibitem[Scicluna and Vella-Zarb(2020)]{scicluna2020evolution}
M.~C. Scicluna and L.~Vella-Zarb, \emph{ACS Applied Nano Materials}, 2020, \textbf{3}, 3097–3115\relax
\mciteBstWouldAddEndPuncttrue
\mciteSetBstMidEndSepPunct{\mcitedefaultmidpunct}
{\mcitedefaultendpunct}{\mcitedefaultseppunct}\relax
\EndOfBibitem
\bibitem[Guo \emph{et~al.}(2022)Guo, Liu, Wu, Sun, and Yang]{guo2022covalent}
H.~Guo, Y.~Liu, N.~Wu, L.~Sun and W.~Yang, \emph{ChemistrySelect}, 2022, \textbf{7}, e202202538\relax
\mciteBstWouldAddEndPuncttrue
\mciteSetBstMidEndSepPunct{\mcitedefaultmidpunct}
{\mcitedefaultendpunct}{\mcitedefaultseppunct}\relax
\EndOfBibitem
\bibitem[Wang \emph{et~al.}(2017)Wang, Dong, Ge, Jiang, and Xu]{wang2017fluorene}
L.~Wang, B.~Dong, R.~Ge, F.~Jiang and J.~Xu, \emph{ACS Applied Materials \& Interfaces}, 2017, \textbf{9}, 7108–7114\relax
\mciteBstWouldAddEndPuncttrue
\mciteSetBstMidEndSepPunct{\mcitedefaultmidpunct}
{\mcitedefaultendpunct}{\mcitedefaultseppunct}\relax
\EndOfBibitem
\bibitem[Chumakov \emph{et~al.}(2016)Chumakov, Aksakal, Dimoglo, Ata, and Palomares-Sánchez]{chumakov2016first}
Y.~Chumakov, F.~Aksakal, A.~Dimoglo, A.~Ata and S.~A. Palomares-Sánchez, \emph{Journal of Electronic Materials}, 2016, \textbf{45}, 3445–3452\relax
\mciteBstWouldAddEndPuncttrue
\mciteSetBstMidEndSepPunct{\mcitedefaultmidpunct}
{\mcitedefaultendpunct}{\mcitedefaultseppunct}\relax
\EndOfBibitem
\bibitem[Wang \emph{et~al.}(2020)Wang, Xu, Yue, Yu, Shui, Huang, and Chen]{wang2020semiconductive}
S.~Wang, X.~Xu, Y.~Yue, K.~Yu, Q.~Shui, N.~Huang and H.~Chen, \emph{Small Structures}, 2020, \textbf{1}, 2000021\relax
\mciteBstWouldAddEndPuncttrue
\mciteSetBstMidEndSepPunct{\mcitedefaultmidpunct}
{\mcitedefaultendpunct}{\mcitedefaultseppunct}\relax
\EndOfBibitem
\bibitem[Fan \emph{et~al.}(2018)Fan, Mundstock, Feldhoff, Knebel, Gu, Meng, and Caro]{fan2018covalent}
H.~Fan, A.~Mundstock, A.~Feldhoff, A.~Knebel, J.~Gu, H.~Meng and J.~Caro, \emph{Journal of the American Chemical Society}, 2018, \textbf{140}, 10094–10098\relax
\mciteBstWouldAddEndPuncttrue
\mciteSetBstMidEndSepPunct{\mcitedefaultmidpunct}
{\mcitedefaultendpunct}{\mcitedefaultseppunct}\relax
\EndOfBibitem
\bibitem[Wu and Yang(2017)]{wu2017applications}
M.-X. Wu and Y.-W. Yang, \emph{Chinese Chemical Letters}, 2017, \textbf{28}, 1135–1143\relax
\mciteBstWouldAddEndPuncttrue
\mciteSetBstMidEndSepPunct{\mcitedefaultmidpunct}
{\mcitedefaultendpunct}{\mcitedefaultseppunct}\relax
\EndOfBibitem
\bibitem[Zhou \emph{et~al.}(2018)Zhou, Yan, Lu, Zhu, Han, Chen, and Ren]{zhou2018routes}
X.~Zhou, Y.~Yan, X.~Lu, H.~Zhu, X.~Han, G.~Chen and Z.~Ren, \emph{Materials Today}, 2018, \textbf{21}, 974–988\relax
\mciteBstWouldAddEndPuncttrue
\mciteSetBstMidEndSepPunct{\mcitedefaultmidpunct}
{\mcitedefaultendpunct}{\mcitedefaultseppunct}\relax
\EndOfBibitem
\bibitem[Evans \emph{et~al.}(2021)Evans, Ryder, Ji, Strauss, Corcos, Vitaku, Flanders, Bisbey, and Dichtel]{evans2021trends}
A.~M. Evans, M.~R. Ryder, W.~Ji, M.~J. Strauss, A.~R. Corcos, E.~Vitaku, N.~C. Flanders, R.~P. Bisbey and W.~R. Dichtel, \emph{Faraday Discussions}, 2021, \textbf{225}, 226–240\relax
\mciteBstWouldAddEndPuncttrue
\mciteSetBstMidEndSepPunct{\mcitedefaultmidpunct}
{\mcitedefaultendpunct}{\mcitedefaultseppunct}\relax
\EndOfBibitem
\bibitem[Zou \emph{et~al.}(2013)Zou, Ren, and Zhu]{zou2013topology}
X.~Zou, H.~Ren and G.~Zhu, \emph{Chemical Communications}, 2013, \textbf{49}, 3925\relax
\mciteBstWouldAddEndPuncttrue
\mciteSetBstMidEndSepPunct{\mcitedefaultmidpunct}
{\mcitedefaultendpunct}{\mcitedefaultseppunct}\relax
\EndOfBibitem
\bibitem[Pang \emph{et~al.}(2017)Pang, Zhou, Liang, Qi, and Zhao]{pang2017regulating}
Z.-F. Pang, T.-Y. Zhou, R.-R. Liang, Q.-Y. Qi and X.~Zhao, \emph{Chemical Science}, 2017, \textbf{8}, 3866–3870\relax
\mciteBstWouldAddEndPuncttrue
\mciteSetBstMidEndSepPunct{\mcitedefaultmidpunct}
{\mcitedefaultendpunct}{\mcitedefaultseppunct}\relax
\EndOfBibitem
\bibitem[Gui \emph{et~al.}(2020)Gui, Lin, Ding, Gao, Mal, and Wang]{gui2020three}
B.~Gui, G.~Lin, H.~Ding, C.~Gao, A.~Mal and C.~Wang, \emph{Accounts of Chemical Research}, 2020, \textbf{53}, 2225–2234\relax
\mciteBstWouldAddEndPuncttrue
\mciteSetBstMidEndSepPunct{\mcitedefaultmidpunct}
{\mcitedefaultendpunct}{\mcitedefaultseppunct}\relax
\EndOfBibitem
\bibitem[Xu \emph{et~al.}(2018)Xu, Chen, Zhou, and Li]{xu2018thermal}
X.~Xu, J.~Chen, J.~Zhou and B.~Li, \emph{Advanced Materials}, 2018, \textbf{30}, 1705544\relax
\mciteBstWouldAddEndPuncttrue
\mciteSetBstMidEndSepPunct{\mcitedefaultmidpunct}
{\mcitedefaultendpunct}{\mcitedefaultseppunct}\relax
\EndOfBibitem
\bibitem[Choy(1977)]{choy1977thermal}
C.~Choy, \emph{Polymer}, 1977, \textbf{18}, 984–1004\relax
\mciteBstWouldAddEndPuncttrue
\mciteSetBstMidEndSepPunct{\mcitedefaultmidpunct}
{\mcitedefaultendpunct}{\mcitedefaultseppunct}\relax
\EndOfBibitem
\bibitem[Utimula \emph{et~al.}(2019)Utimula, Ichibha, Maezono, and Hongo]{utimula2019ab}
K.~Utimula, T.~Ichibha, R.~Maezono and K.~Hongo, \emph{Chemistry of Materials}, 2019, \textbf{31}, 4649–4656\relax
\mciteBstWouldAddEndPuncttrue
\mciteSetBstMidEndSepPunct{\mcitedefaultmidpunct}
{\mcitedefaultendpunct}{\mcitedefaultseppunct}\relax
\EndOfBibitem
\bibitem[Lindsay \emph{et~al.}(2012)Lindsay, Broido, and Reinecke]{lindsay2012thermal}
L.~Lindsay, D.~A. Broido and T.~L. Reinecke, \emph{Phys. Rev. Lett.}, 2012, \textbf{109}, 095901\relax
\mciteBstWouldAddEndPuncttrue
\mciteSetBstMidEndSepPunct{\mcitedefaultmidpunct}
{\mcitedefaultendpunct}{\mcitedefaultseppunct}\relax
\EndOfBibitem
\bibitem[Romero \emph{et~al.}(2015)Romero, Gross, Verstraete, and Hellman]{romero2015thermal}
A.~H. Romero, E.~K.~U. Gross, M.~J. Verstraete and O.~Hellman, \emph{Phys. Rev. B}, 2015, \textbf{91}, 214310\relax
\mciteBstWouldAddEndPuncttrue
\mciteSetBstMidEndSepPunct{\mcitedefaultmidpunct}
{\mcitedefaultendpunct}{\mcitedefaultseppunct}\relax
\EndOfBibitem
\bibitem[Li and Brédas(2021)]{li2021impact}
H.~Li and J.-L. Brédas, \emph{Chemistry of Materials}, 2021, \textbf{33}, 4529–4540\relax
\mciteBstWouldAddEndPuncttrue
\mciteSetBstMidEndSepPunct{\mcitedefaultmidpunct}
{\mcitedefaultendpunct}{\mcitedefaultseppunct}\relax
\EndOfBibitem
\bibitem[Li \emph{et~al.}(2020)Li, Liu, Li, Wang, Zhao, Ding, Feng, and Han]{li2020defective}
Z.~Li, Z.~Liu, Z.~Li, T.~Wang, F.~Zhao, X.~Ding, W.~Feng and B.~Han, \emph{Advanced Functional Materials}, 2020, \textbf{30}, 1909267\relax
\mciteBstWouldAddEndPuncttrue
\mciteSetBstMidEndSepPunct{\mcitedefaultmidpunct}
{\mcitedefaultendpunct}{\mcitedefaultseppunct}\relax
\EndOfBibitem
\bibitem[Rapaport(2004)]{rapaport2004art}
D.~C. Rapaport, \emph{The Art of Molecular Dynamics Simulation}, Cambridge University Press, 2004\relax
\mciteBstWouldAddEndPuncttrue
\mciteSetBstMidEndSepPunct{\mcitedefaultmidpunct}
{\mcitedefaultendpunct}{\mcitedefaultseppunct}\relax
\EndOfBibitem
\bibitem[Islamov \emph{et~al.}(2023)Islamov, Babaei, Anderson, Sezginel, Long, McGaughey, Gomez-Gualdron, and Wilmer]{islamov2023high}
M.~Islamov, H.~Babaei, R.~Anderson, K.~B. Sezginel, J.~R. Long, A.~J.~H. McGaughey, D.~A. Gomez-Gualdron and C.~E. Wilmer, \emph{npj Computational Materials}, 2023, \textbf{9}, DOI: 10.1038/s41524--022--00961--x\relax
\mciteBstWouldAddEndPuncttrue
\mciteSetBstMidEndSepPunct{\mcitedefaultmidpunct}
{\mcitedefaultendpunct}{\mcitedefaultseppunct}\relax
\EndOfBibitem
\bibitem[Thakur and Giri(2024)]{thakur2024pushing}
S.~Thakur and A.~Giri, \emph{Small}, 2024, \textbf{20}, 2401702\relax
\mciteBstWouldAddEndPuncttrue
\mciteSetBstMidEndSepPunct{\mcitedefaultmidpunct}
{\mcitedefaultendpunct}{\mcitedefaultseppunct}\relax
\EndOfBibitem
\bibitem[Huang \emph{et~al.}(2023)Huang, Ma, Zhao, Wang, and Ju]{huang2023exploring}
X.~Huang, S.~Ma, C.~Y. Zhao, H.~Wang and S.~Ju, \emph{npj Computational Materials}, 2023, \textbf{9}, DOI: 10.1038/s41524--023--01154--w\relax
\mciteBstWouldAddEndPuncttrue
\mciteSetBstMidEndSepPunct{\mcitedefaultmidpunct}
{\mcitedefaultendpunct}{\mcitedefaultseppunct}\relax
\EndOfBibitem
\bibitem[Ma \emph{et~al.}(2022)Ma, Zhang, and Luo]{ma2022exploring}
R.~Ma, H.~Zhang and T.~Luo, \emph{ACS Applied Materials \& Interfaces}, 2022, \textbf{14}, 15587--15598\relax
\mciteBstWouldAddEndPuncttrue
\mciteSetBstMidEndSepPunct{\mcitedefaultmidpunct}
{\mcitedefaultendpunct}{\mcitedefaultseppunct}\relax
\EndOfBibitem
\bibitem[Wu \emph{et~al.}(2019)Wu, Kondo, Kakimoto, Yang, Yamada, Kuwajima, Lambard, Hongo, Xu, Shiomi,\emph{et~al.}]{wu2019machine}
S.~Wu, Y.~Kondo, M.-a. Kakimoto, B.~Yang, H.~Yamada, I.~Kuwajima, G.~Lambard, K.~Hongo, Y.~Xu, J.~Shiomi \emph{et~al.}, \emph{Npj Computational Materials}, 2019, \textbf{5}, 66\relax
\mciteBstWouldAddEndPuncttrue
\mciteSetBstMidEndSepPunct{\mcitedefaultmidpunct}
{\mcitedefaultendpunct}{\mcitedefaultseppunct}\relax
\EndOfBibitem
\bibitem[Ishikiriyama(2022)]{ishikiriyama2022polymer}
K.~Ishikiriyama, \emph{Thermochimica Acta}, 2022, \textbf{708}, 179135\relax
\mciteBstWouldAddEndPuncttrue
\mciteSetBstMidEndSepPunct{\mcitedefaultmidpunct}
{\mcitedefaultendpunct}{\mcitedefaultseppunct}\relax
\EndOfBibitem
\bibitem[Xu and Luo(2024)]{xu2024unlocking}
J.~Xu and T.~Luo, \emph{npj Computational Materials}, 2024, \textbf{10}, 74\relax
\mciteBstWouldAddEndPuncttrue
\mciteSetBstMidEndSepPunct{\mcitedefaultmidpunct}
{\mcitedefaultendpunct}{\mcitedefaultseppunct}\relax
\EndOfBibitem
\bibitem[Hu \emph{et~al.}(2024)Hu, Wang, and Ma]{hu2024thermally}
Y.~Hu, Q.~Wang and H.~Ma, \emph{Journal of Applied Physics}, 2024, \textbf{135}, 120701\relax
\mciteBstWouldAddEndPuncttrue
\mciteSetBstMidEndSepPunct{\mcitedefaultmidpunct}
{\mcitedefaultendpunct}{\mcitedefaultseppunct}\relax
\EndOfBibitem
\bibitem[Yang \emph{et~al.}(2021)Yang, Zhang, Lai, Wang, Yang, and Yu]{yang2021accelerating}
P.~Yang, H.~Zhang, X.~Lai, K.~Wang, Q.~Yang and D.~Yu, \emph{ACS Omega}, 2021, \textbf{6}, 17149–17161\relax
\mciteBstWouldAddEndPuncttrue
\mciteSetBstMidEndSepPunct{\mcitedefaultmidpunct}
{\mcitedefaultendpunct}{\mcitedefaultseppunct}\relax
\EndOfBibitem
\bibitem[Cao \emph{et~al.}(2022)Cao, Zhang, He, Xue, Huang, and Zhong]{cao2022machine}
X.~Cao, Z.~Zhang, Y.~He, W.~Xue, H.~Huang and C.~Zhong, \emph{Industrial \& Engineering Chemistry Research}, 2022, \textbf{61}, 11116–11123\relax
\mciteBstWouldAddEndPuncttrue
\mciteSetBstMidEndSepPunct{\mcitedefaultmidpunct}
{\mcitedefaultendpunct}{\mcitedefaultseppunct}\relax
\EndOfBibitem
\bibitem[De~Vos \emph{et~al.}(2024)De~Vos, Ravichandran, Borgmans, Vanduyfhuys, Van Der~Voort, Rogge, and Van~Speybroeck]{de2024high}
J.~S. De~Vos, S.~Ravichandran, S.~Borgmans, L.~Vanduyfhuys, P.~Van Der~Voort, S.~M.~J. Rogge and V.~Van~Speybroeck, \emph{Chemistry of Materials}, 2024, \textbf{36}, 4315–4330\relax
\mciteBstWouldAddEndPuncttrue
\mciteSetBstMidEndSepPunct{\mcitedefaultmidpunct}
{\mcitedefaultendpunct}{\mcitedefaultseppunct}\relax
\EndOfBibitem
\bibitem[Zhao and Chung(2024)]{zhao2024pacman}
G.~Zhao and Y.~G. Chung, \emph{Journal of Chemical Theory and Computation}, 2024, \textbf{20}, 5368–5380\relax
\mciteBstWouldAddEndPuncttrue
\mciteSetBstMidEndSepPunct{\mcitedefaultmidpunct}
{\mcitedefaultendpunct}{\mcitedefaultseppunct}\relax
\EndOfBibitem
\bibitem[Korolev and Mitrofanov(2024)]{korolev2024coarse}
V.~Korolev and A.~Mitrofanov, \emph{Journal of Chemical Information and Modeling}, 2024, \textbf{64}, 1919--1931\relax
\mciteBstWouldAddEndPuncttrue
\mciteSetBstMidEndSepPunct{\mcitedefaultmidpunct}
{\mcitedefaultendpunct}{\mcitedefaultseppunct}\relax
\EndOfBibitem
\bibitem[Kang \emph{et~al.}(2023)Kang, Park, Smit, and Kim]{kang2023multi}
Y.~Kang, H.~Park, B.~Smit and J.~Kim, \emph{Nature Machine Intelligence}, 2023, \textbf{5}, 309–318\relax
\mciteBstWouldAddEndPuncttrue
\mciteSetBstMidEndSepPunct{\mcitedefaultmidpunct}
{\mcitedefaultendpunct}{\mcitedefaultseppunct}\relax
\EndOfBibitem
\bibitem[Park \emph{et~al.}(2023)Park, Kang, and Kim]{park2023enhancing}
H.~Park, Y.~Kang and J.~Kim, \emph{ACS Applied Materials \& Interfaces}, 2023, \textbf{15}, 56375–56385\relax
\mciteBstWouldAddEndPuncttrue
\mciteSetBstMidEndSepPunct{\mcitedefaultmidpunct}
{\mcitedefaultendpunct}{\mcitedefaultseppunct}\relax
\EndOfBibitem
\bibitem[Mercado \emph{et~al.}(2018)Mercado, Fu, Yakutovich, Talirz, Haranczyk, and Smit]{mercado2018silico}
R.~Mercado, R.-S. Fu, A.~V. Yakutovich, L.~Talirz, M.~Haranczyk and B.~Smit, \emph{Chemistry of Materials}, 2018, \textbf{30}, 5069–5086\relax
\mciteBstWouldAddEndPuncttrue
\mciteSetBstMidEndSepPunct{\mcitedefaultmidpunct}
{\mcitedefaultendpunct}{\mcitedefaultseppunct}\relax
\EndOfBibitem
\bibitem[Freitas \emph{et~al.}(2017)Freitas, Borges, Merlini, Barra, and Esteves]{freitas2017thermal}
S.~K.~S. Freitas, R.~S. Borges, C.~Merlini, G.~M.~O. Barra and P.~M. Esteves, \emph{The Journal of Physical Chemistry C}, 2017, \textbf{121}, 27247–27252\relax
\mciteBstWouldAddEndPuncttrue
\mciteSetBstMidEndSepPunct{\mcitedefaultmidpunct}
{\mcitedefaultendpunct}{\mcitedefaultseppunct}\relax
\EndOfBibitem
\bibitem[Vaswani \emph{et~al.}(2017)Vaswani, Shazeer, Parmar, Uszkoreit, Jones, Gomez, Kaiser, and Polosukhin]{vaswani2017attention}
A.~Vaswani, N.~Shazeer, N.~Parmar, J.~Uszkoreit, L.~Jones, A.~N. Gomez, L.~Kaiser and I.~Polosukhin, \emph{arXiv}, 2017,  preprint, arXiv:1706.03762, \url{https://arxiv.org/abs/1706.03762}\relax
\mciteBstWouldAddEndPuncttrue
\mciteSetBstMidEndSepPunct{\mcitedefaultmidpunct}
{\mcitedefaultendpunct}{\mcitedefaultseppunct}\relax
\EndOfBibitem
\bibitem[Xie and Grossman(2018)]{xie2018crystal}
T.~Xie and J.~C. Grossman, \emph{Phys. Rev. Lett.}, 2018, \textbf{120}, 145301\relax
\mciteBstWouldAddEndPuncttrue
\mciteSetBstMidEndSepPunct{\mcitedefaultmidpunct}
{\mcitedefaultendpunct}{\mcitedefaultseppunct}\relax
\EndOfBibitem
\bibitem[git()]{githubGitHubSangwon91GRIDAY}
\emph{{G}it{H}ub - {S}angwon91/{G}{R}{I}{D}{A}{Y}: {E}nergy shape calculator for the porous materials --- github.com}, \url{https://github.com/Sangwon91/GRIDAY}, [Accessed 14-06-2024]\relax
\mciteBstWouldAddEndPuncttrue
\mciteSetBstMidEndSepPunct{\mcitedefaultmidpunct}
{\mcitedefaultendpunct}{\mcitedefaultseppunct}\relax
\EndOfBibitem
\bibitem[Dosovitskiy \emph{et~al.}(2020)Dosovitskiy, Beyer, Kolesnikov, Weissenborn, Zhai, Unterthiner, Dehghani, Minderer, Heigold, Gelly, Uszkoreit, and Houlsby]{alexey2020image}
A.~Dosovitskiy, L.~Beyer, A.~Kolesnikov, D.~Weissenborn, X.~Zhai, T.~Unterthiner, M.~Dehghani, M.~Minderer, G.~Heigold, S.~Gelly, J.~Uszkoreit and N.~Houlsby, \emph{arXiv}, 2020,  prepint, arXiv:2010.11929v2, \url{https://arxiv.org/abs/2010.11929}\relax
\mciteBstWouldAddEndPuncttrue
\mciteSetBstMidEndSepPunct{\mcitedefaultmidpunct}
{\mcitedefaultendpunct}{\mcitedefaultseppunct}\relax
\EndOfBibitem
\bibitem[Luo \emph{et~al.}(2017)Luo, Huang, and Huang]{luo2018decreased}
D.~Luo, C.~Huang and Z.~Huang, \emph{Journal of Heat Transfer}, 2017, \textbf{140}, 031302\relax
\mciteBstWouldAddEndPuncttrue
\mciteSetBstMidEndSepPunct{\mcitedefaultmidpunct}
{\mcitedefaultendpunct}{\mcitedefaultseppunct}\relax
\EndOfBibitem
\bibitem[Feng \emph{et~al.}(2020)Feng, He, Rai, Hun, Liu, and Shrestha]{feng2020size}
T.~Feng, J.~He, A.~Rai, D.~Hun, J.~Liu and S.~S. Shrestha, \emph{Phys. Rev. Appl.}, 2020, \textbf{14}, 044023\relax
\mciteBstWouldAddEndPuncttrue
\mciteSetBstMidEndSepPunct{\mcitedefaultmidpunct}
{\mcitedefaultendpunct}{\mcitedefaultseppunct}\relax
\EndOfBibitem
\bibitem[Feng \emph{et~al.}(2020)Feng, Feng, Liu, Zhang, Yan, and Zhang]{feng2020thermal}
D.~Feng, Y.~Feng, Y.~Liu, W.~Zhang, Y.~Yan and X.~Zhang, \emph{The Journal of Physical Chemistry C}, 2020, \textbf{124}, 8386–8393\relax
\mciteBstWouldAddEndPuncttrue
\mciteSetBstMidEndSepPunct{\mcitedefaultmidpunct}
{\mcitedefaultendpunct}{\mcitedefaultseppunct}\relax
\EndOfBibitem
\bibitem[Ying \emph{et~al.}(2021)Ying, Zhang, and Zhong]{ying2021effect}
P.~Ying, J.~Zhang and Z.~Zhong, \emph{The Journal of Physical Chemistry C}, 2021, \textbf{125}, 12991–13001\relax
\mciteBstWouldAddEndPuncttrue
\mciteSetBstMidEndSepPunct{\mcitedefaultmidpunct}
{\mcitedefaultendpunct}{\mcitedefaultseppunct}\relax
\EndOfBibitem
\bibitem[Wang \emph{et~al.}(2023)Wang, Ren, Han, Zhou, Wong, Bai, Sun, and Zeng]{wang2023molecular}
S.~Wang, L.~Ren, M.~Han, W.~Zhou, C.~Wong, X.~Bai, R.~Sun and X.~Zeng, \emph{Nanoscale}, 2023, \textbf{15}, 8706–8715\relax
\mciteBstWouldAddEndPuncttrue
\mciteSetBstMidEndSepPunct{\mcitedefaultmidpunct}
{\mcitedefaultendpunct}{\mcitedefaultseppunct}\relax
\EndOfBibitem
\bibitem[Ying \emph{et~al.}(2020)Ying, Zhang, Zhang, and Zhong]{ying2020impacts}
P.~Ying, J.~Zhang, X.~Zhang and Z.~Zhong, \emph{The Journal of Physical Chemistry C}, 2020, \textbf{124}, 6274–6283\relax
\mciteBstWouldAddEndPuncttrue
\mciteSetBstMidEndSepPunct{\mcitedefaultmidpunct}
{\mcitedefaultendpunct}{\mcitedefaultseppunct}\relax
\EndOfBibitem
\bibitem[Breiman(2001)]{breiman2001random}
L.~Breiman, \emph{Machine Learning}, 2001, \textbf{45}, 5–32\relax
\mciteBstWouldAddEndPuncttrue
\mciteSetBstMidEndSepPunct{\mcitedefaultmidpunct}
{\mcitedefaultendpunct}{\mcitedefaultseppunct}\relax
\EndOfBibitem
\bibitem[Friedman(2001)]{friedman2001greedy}
J.~H. Friedman, \emph{The Annals of Statistics}, 2001, \textbf{29}, 1189–1232\relax
\mciteBstWouldAddEndPuncttrue
\mciteSetBstMidEndSepPunct{\mcitedefaultmidpunct}
{\mcitedefaultendpunct}{\mcitedefaultseppunct}\relax
\EndOfBibitem
\bibitem[Chen and Guestrin(2016)]{chen2016xgboost}
T.~Chen and C.~Guestrin, Proceedings of the 22nd ACM SIGKDD International Conference on Knowledge Discovery and Data Mining, 2016, p. 785–794\relax
\mciteBstWouldAddEndPuncttrue
\mciteSetBstMidEndSepPunct{\mcitedefaultmidpunct}
{\mcitedefaultendpunct}{\mcitedefaultseppunct}\relax
\EndOfBibitem
\bibitem[Freund and Schapire(1997)]{freund1995desicion}
Y.~Freund and R.~E. Schapire, \emph{Journal of Computer and System Sciences}, 1997, \textbf{55}, 119–139\relax
\mciteBstWouldAddEndPuncttrue
\mciteSetBstMidEndSepPunct{\mcitedefaultmidpunct}
{\mcitedefaultendpunct}{\mcitedefaultseppunct}\relax
\EndOfBibitem
\bibitem[Pedregosa \emph{et~al.}(2011)Pedregosa, Varoquaux, Gramfort, Michel, Thirion, Grisel, Blondel, Prettenhofer, Weiss, Dubourg, Vanderplas, Passos, Cournapeau, Brucher, Perrot, and Duchesnay]{scikit-learn}
F.~Pedregosa, G.~Varoquaux, A.~Gramfort, V.~Michel, B.~Thirion, O.~Grisel, M.~Blondel, P.~Prettenhofer, R.~Weiss, V.~Dubourg, J.~Vanderplas, A.~Passos, D.~Cournapeau, M.~Brucher, M.~Perrot and E.~Duchesnay, \emph{Journal of Machine Learning Research}, 2011, \textbf{12}, 2825--2830\relax
\mciteBstWouldAddEndPuncttrue
\mciteSetBstMidEndSepPunct{\mcitedefaultmidpunct}
{\mcitedefaultendpunct}{\mcitedefaultseppunct}\relax
\EndOfBibitem
\bibitem[Lundberg(2017)]{lundberg2017shap}
S.~Lundberg, \emph{arXiv}, 2017,  preprint, arXiv:1705.07874v2, \url{https://arxiv.org/abs/1705.07874}\relax
\mciteBstWouldAddEndPuncttrue
\mciteSetBstMidEndSepPunct{\mcitedefaultmidpunct}
{\mcitedefaultendpunct}{\mcitedefaultseppunct}\relax
\EndOfBibitem
\bibitem[Lan \emph{et~al.}(2018)Lan, Han, Tong, Huang, Yang, Liu, Zhao, and Zhong]{lan2018materials}
Y.~Lan, X.~Han, M.~Tong, H.~Huang, Q.~Yang, D.~Liu, X.~Zhao and C.~Zhong, \emph{Nature Communications}, 2018, \textbf{9}, 5274\relax
\mciteBstWouldAddEndPuncttrue
\mciteSetBstMidEndSepPunct{\mcitedefaultmidpunct}
{\mcitedefaultendpunct}{\mcitedefaultseppunct}\relax
\EndOfBibitem
\end{mcitethebibliography}


\begin{thebibliography}{10}

\bibitem{githubGitHubSangwon91GRIDAY}
{G}it{H}ub - {S}angwon91/{G}{R}{I}{D}{A}{Y}: {E}nergy shape calculator for the porous materials --- github.com.
\newblock \url{https://github.com/Sangwon91/GRIDAY}.
\newblock [Accessed 14-06-2024].

\bibitem{boyd2017force}
P.~G. Boyd, S.~M. Moosavi, M.~Witman, and B.~Smit.
\newblock Force-field prediction of materials properties in metal-organic frameworks.
\newblock {\em The Journal of Physical Chemistry Letters}, 8(2):357–363, Jan. 2017.

\bibitem{breiman2001random}
L.~Breiman.
\newblock Random forests.
\newblock {\em Machine Learning}, 45(1):5–32, 2001.

\bibitem{chen2016xgboost}
T.~Chen and C.~Guestrin.
\newblock Xgboost: A scalable tree boosting system.
\newblock In {\em Proceedings of the 22nd ACM SIGKDD International Conference on Knowledge Discovery and Data Mining}, volume~11 of {\em KDD ’16}, page 785–794. ACM, August 2016.

\bibitem{daliran2024defects}
S.~Daliran, M.~Blanco, A.~Dhakshinamoorthy, A.~R. Oveisi, J.~Alem{\'a}n, and H.~Garc{\'\i}a.
\newblock Defects and disorder in covalent organic frameworks for advanced applications.
\newblock {\em Advanced Functional Materials}, 34(18):2312912, 2024.

\bibitem{daliran2024probing}
S.~Daliran, A.~R. Oveisi, A.~Dhakshinamoorthy, and H.~Garcia.
\newblock Probing defects in covalent organic frameworks.
\newblock {\em ACS Applied Materials \& Interfaces}, 16(38):50096--50114, 2024.

\bibitem{evans2021thermally}
A.~M. Evans, A.~Giri, V.~K. Sangwan, S.~Xun, M.~Bartnof, C.~G. Torres-Castanedo, H.~B. Balch, M.~S. Rahn, N.~P. Bradshaw, E.~Vitaku, D.~W. Burke, H.~Li, M.~J. Bedzyk, F.~Wang, J.-L. Brédas, J.~A. Malen, A.~J.~H. McGaughey, M.~C. Hersam, W.~R. Dichtel, and P.~E. Hopkins.
\newblock Thermally conductive ultra-low-k dielectric layers based on two-dimensional covalent organic frameworks.
\newblock {\em Nature Materials}, 20(8):1142–1148, Mar. 2021.

\bibitem{freund1995desicion}
Y.~Freund and R.~E. Schapire.
\newblock A decision-theoretic generalization of on-line learning and an application to boosting.
\newblock {\em Journal of Computer and System Sciences}, 55(1):119–139, August 1997.

\bibitem{friedman2001greedy}
J.~H. Friedman.
\newblock Greedy function approximation: A gradient boosting machine.
\newblock {\em The Annals of Statistics}, 29(5):1189–1232, October 2001.

\bibitem{hirel2015atomsk}
P.~Hirel.
\newblock Atomsk: A tool for manipulating and converting atomic data files.
\newblock {\em Computer Physics Communications}, 197:212–219, Dec. 2015.

\bibitem{loshchilov2017decoupled}
I.~Loshchilov and F.~Hutter.
\newblock Decoupled weight decay regularization, 2017.

\bibitem{mayo1990dreiding}
S.~L. Mayo, B.~D. Olafson, and W.~A. Goddard.
\newblock Dreiding: a generic force field for molecular simulations.
\newblock {\em Journal of Physical chemistry}, 94(26):8897--8909, 1990.

\bibitem{mercado2018silico}
R.~Mercado, R.-S. Fu, A.~V. Yakutovich, L.~Talirz, M.~Haranczyk, and B.~Smit.
\newblock In silico design of 2d and 3d covalent organic frameworks for methane storage applications.
\newblock {\em Chemistry of Materials}, 30(15):5069–5086, June 2018.

\bibitem{scikit-learn}
F.~Pedregosa, G.~Varoquaux, A.~Gramfort, V.~Michel, B.~Thirion, O.~Grisel, M.~Blondel, P.~Prettenhofer, R.~Weiss, V.~Dubourg, J.~Vanderplas, A.~Passos, D.~Cournapeau, M.~Brucher, M.~Perrot, and E.~Duchesnay.
\newblock Scikit-learn: Machine learning in {P}ython.
\newblock {\em Journal of Machine Learning Research}, 12:2825--2830, 2011.

\bibitem{stuart2000reactive}
S.~J. Stuart, A.~B. Tutein, and J.~A. Harrison.
\newblock A reactive potential for hydrocarbons with intermolecular interactions.
\newblock {\em The Journal of chemical physics}, 112(14):6472--6486, 2000.

\bibitem{thakur2024pushing}
S.~Thakur and A.~Giri.
\newblock Pushing the limits of heat conduction in covalent organic frameworks through high‐throughput screening of their thermal conductivity.
\newblock {\em Small}, 20(32):2401702, Apr. 2024.

\bibitem{thomas2010predicting}
J.~A. Thomas, J.~E. Turney, R.~M. Iutzi, C.~H. Amon, and A.~J. McGaughey.
\newblock Predicting phonon dispersion relations and lifetimes from the spectral energy density.
\newblock {\em Physical Review B—Condensed Matter and Materials Physics}, 81(8):081411, 2010.

\bibitem{thompson2022lammps}
A.~P. Thompson, H.~M. Aktulga, R.~Berger, D.~S. Bolintineanu, W.~M. Brown, P.~S. Crozier, P.~J. in~’t Veld, A.~Kohlmeyer, S.~G. Moore, T.~D. Nguyen, R.~Shan, M.~J. Stevens, J.~Tranchida, C.~Trott, and S.~J. Plimpton.
\newblock Lammps - a flexible simulation tool for particle-based materials modeling at the atomic, meso, and continuum scales.
\newblock {\em Computer Physics Communications}, 271:108171, Feb. 2022.

\bibitem{wirnsberger2015enhanced}
P.~Wirnsberger, D.~Frenkel, and C.~Dellago.
\newblock An enhanced version of the heat exchange algorithm with excellent energy conservation properties.
\newblock {\em The Journal of Chemical Physics}, 143(12), Sept. 2015.

\bibitem{yu2024unraveling}
A.~Yu, W.~Liu, W.~Xi, M.~Mu, and L.~Shi.
\newblock Unraveling the rapid proton transport mechanism of covalent organic frameworks.
\newblock {\em Chemistry of Materials}, 36(4):1880--1890, 2024.

\end{thebibliography}
\bibliographystyle{rsc} %
\end{document}


\setcounter{figure}{0}
\renewcommand{\figurename}{Fig.}
\renewcommand{\thefigure}{S\arabic{figure}}

\setcounter{table}{0}
\renewcommand{\tablename}{Table}
\renewcommand{\thetable}{S\arabic{table}}

\title{Deep learning reveals key predictors of thermal conductivity in covalent organic frameworks
\\[0.3cm] \Large\textit{Supplementary Information}}

\date{}

\author{Prakash Thakolkaran\textsuperscript{1,$\ddag$}, Yiwen Zheng\textsuperscript{2,$\ddag$}, Yaqi Guo\textsuperscript{1}, Aniruddh Vashisth\textsuperscript{2,\P}, and Siddhant Kumar\textsuperscript{1,\P}}

\maketitle
\newcommand{\rev}[1]{{{#1}}}
\newcommand{\SK}[1]{{{#1}}}
\newcommand{\AV}[1]{{{#1}}}
\newcommand{\PT}[1]{{{#1}}}
\newcommand{\YG}[1]{{{#1}}}
\newcommand{\YZ}[1]{{{#1}}}

\textsuperscript{1}\textit{Department of Materials Science and Engineering, Delft University of Technology, 2628 CD Delft, The Netherlands}\\
\textsuperscript{2}\textit{Department of Mechanical Engineering, University of Washington, Seattle, WA, USA}\\
\textsuperscript{$\ddag$}\textit{Equal contribution}\\
\textsuperscript{\P}\textit{Email: vashisth@uw.edu, sid.kumar@tudelft.nl; equal contribution}

\section{Data generation}

\begin{figure}
\centering
\includegraphics[width=0.7\linewidth]{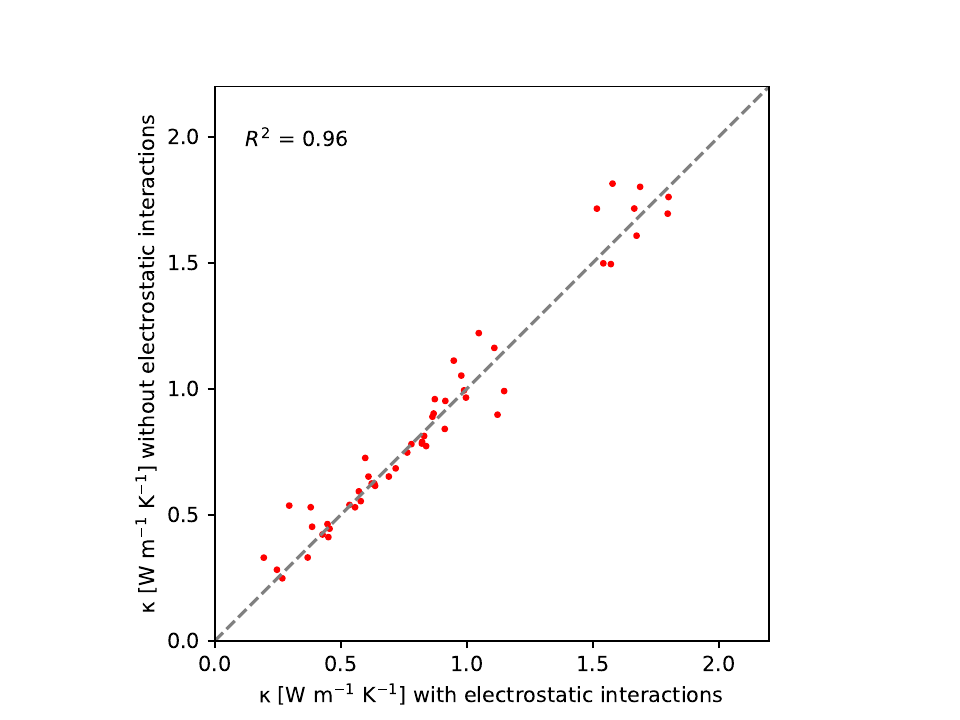}
\caption{{Comparison of thermal conductivity of 50 COFs by MD simulations without electrostatic interactions vs. with electrostatic interactions.}}
\label{fig:charge-layer}
\end{figure}

\begin{figure}
\centering
\includegraphics[width=\linewidth]{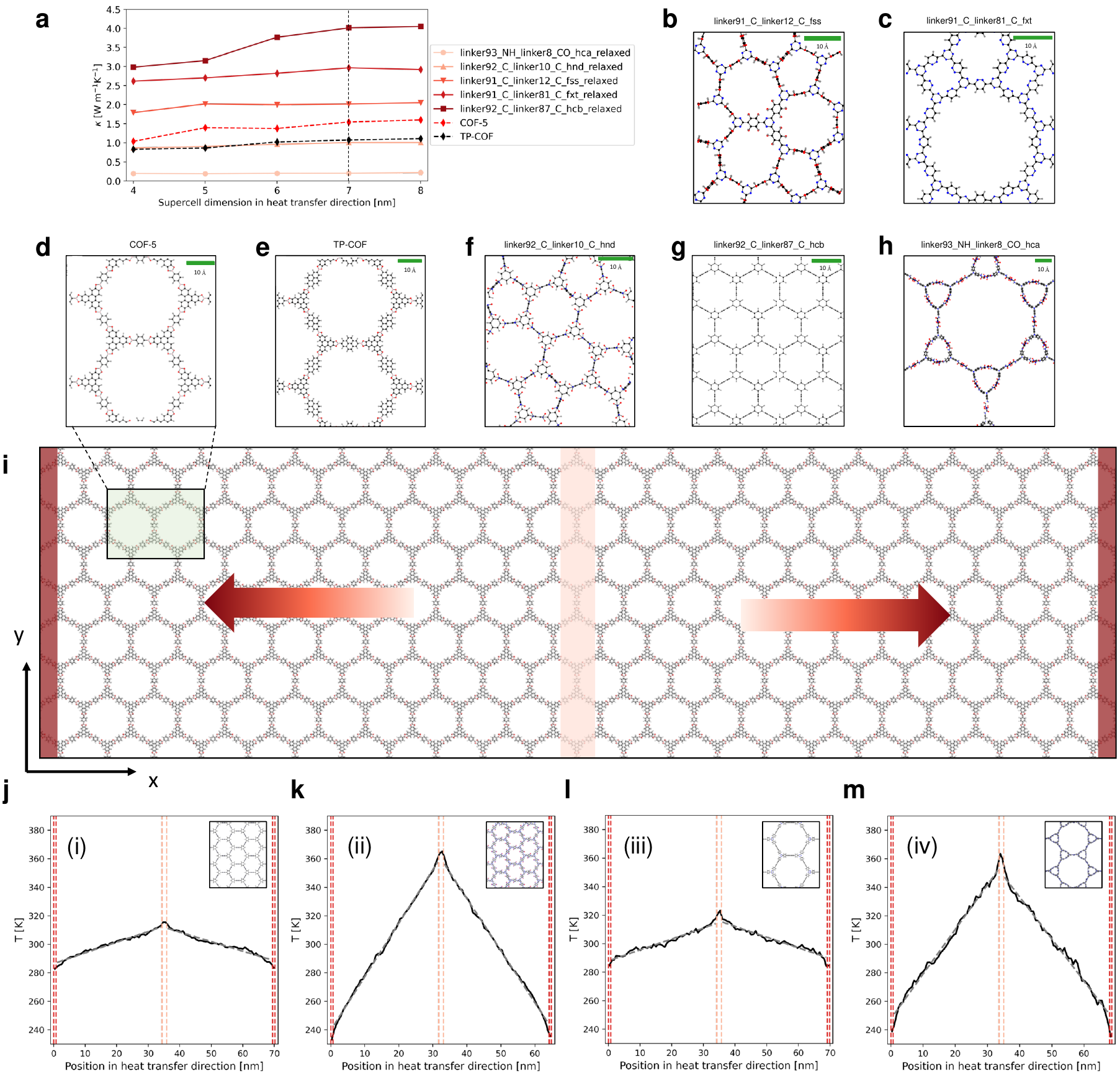}
\caption{Schematic of calculation of in-plane thermal conductivities of COFs by NEMD simulations. (a) Average thermal conductivity $\kappa$ as a function of supercell dimension in heat transfer direction. (b-h) Orthogonal unit cells of the COFs used in the convergence study. (i) COF-5 supercell created by repeating the unit cell in x and y directions. The supercell is divided into 100 bins. The middle two bins are set as heat sources while the two bins nearest to the boundary are set as cold sinks.  (j-m) The average temperature of each bin for each example COF in \figurename~2 of the main article. The spatial temperature distribution is fitted using linear regressions (grey dashed lines) to obtain temperature gradients.}
\label{fig:MD}
\end{figure}

\subsection{Molecular Dynamics}

We carried out MD simulations on two-dimensional COF structures sampled from the unlabeled dataset by Mercado et al. \cite{mercado2018silico} to calculate their in-plane thermal conductivities. The simulations were performed by Large-scale Atomic/Molecular Massively Parallel Simulator (LAMMPS) \cite{thompson2022lammps}. We employed the Dreiding force field \cite{mayo1990dreiding} to describe the interatomic interactions due to its applicability to a wide range of 2D COFs with various chemistries and topologies, as well as compatibility with the unit cell structures from the Mercado dataset \cite{mercado2018silico} which were minimized using Dreiding. {The computational efficiency and broad applicability to organic materials further make Dreiding a suitable choice for generating a comprehensive dataset of thermal conductivity.} Other force fields such as reactive potentials fall short in accommodating more than 2,400 COFs in this study due to limited generalizability and high computational cost \cite{evans2021thermally, yu2024unraveling}. {Future research should focus on developing a novel force field specifically tailored for thermal transport simulations of COFs.} The electrostatic interactions between atoms were ignored due to the high computational cost {and minimal effect on calculated thermal conductivity of COFs, which is confirmed by the good agreement between the calculated thermal conductivity with and without electrostatic interactions (\figurename~\ref{fig:charge-layer}a).}

The COF structures in the Crystallographic Information File (CIF) format extracted from the dataset were converted into orthogonal unit cells by Atomsk \cite{hirel2015atomsk}. We calculated the thermal conductivity in two orthogonal in-plane directions (denoted as x and y) of COFs. To calculate $\kappa_\mathrm{x}$ (e.g., zigzag direction of honeycomb COFs), we repeated each orthogonal unit cell to a supercell of 7 nm by 2 nm in x and y directions, respectively (\figurename~\ref{fig:MD}i). Another supercell of 2 nm by 7 nm was created to calculate $\kappa_\mathrm{y}$ (e.g., armchair direction of honeycomb COFs). Each supercell contained one layer of 2D COF in the z (cross-plane) direction. Unit cell repetition and atom type assignment were done by the LAMMPS Interface package \cite{boyd2017force}. The initial configuration was minimized using the conjugate gradient method and equilibrated by consecutive runs under NPT and NVT ensembles. More specifically, the minimized structure was heated to 300 K in 60 K steps. Each heating step took 100 ps under NVT followed by a 100-ps holding under NPT to relax the structure at constant temperature. The total equilibration time was 1 ns. {All simulations utilized a time step size of 1 fs. For NVT simulations, a Nose-Hoover thermostat with a 0.1 ps damping parameter was employed, while NPT simulations used a Nose-Hoover barostat with a 1 ps damping parameter.} During equilibration, the crystal structures of some COFs became unstable, leading to significant changes (mostly contraction) in the dimensions of the supercells. We filtered 2,471 intact COFs after equilibration by selecting those whose relative change in supercell dimension is under 10\%.

We employed the non-equilibrium molecular dynamics (NEMD) method to calculate in-plane thermal conductivities of COFs. \rev{In contrast to the equilibrium molecular dynamics (EMD) method, which employs the Green-Kubo relation as seen in previous studies \cite{evans2021thermally, thakur2024pushing}, the NEMD method does not depend on the convergence of autocorrelation functions. This makes it more suitable for the efficient calculation of thermal conductivity for a large number of COFs.} In the NEMD setup, the equilibrated structure was divided into 100 bins with equal width in the direction of heat flow. The leftmost and rightmost bins were defined as cold sinks while the two bins in the middle were defined as heat sources (\figurename~\ref{fig:MD}i). Periodic boundary conditions were applied in all three directions. Equal amount of heat was added to each hot bin and subtracted from each cold bin by the enhanced heat exchange algorithm \cite{wirnsberger2015enhanced} at a rate proportional to the total number of atoms in the system:
\begin{equation} \label{eq_heat}
    \frac{\dd E}{\dd t} = k \cdot n_\mathrm{atom},
\end{equation}
where the constant $k = 10^{-7}~\mathrm{kcal} \cdot \mathrm{mol}^{-1} \cdot \mathrm{fs}^{-1}$. Since thermal conductivity is typically correlated with density, Equation \ref{eq_heat} ensures a reasonable temperature difference between heat sources and cold sinks. This setup was maintained in the NVE ensemble for 2 ns and the temperature of each bin except heat sources and cold sinks was recorded and averaged over the last 1 ns. Two temperature gradients were obtained from two halves of the system and fit to linear regressions (\figurename~\ref{fig:MD}j-m). The thermal conductivity is calculated by Fourier's law:
\begin{equation} \label{eq_fourier}
    \kappa = \frac{1}{S} \frac{\dd E}{\dd t} \dfrac{\dd L}{\dd T},
\end{equation}
where $S$ is the cross-sectional area perpendicular to the direction of heat flow and ${\dd T}/{\dd L}$ is the average absolute slopes of the linear fits from both halves. The same method was used to calculate $\kappa_\mathrm{x}$ and $\kappa_\mathrm{y}$ on a total number of 2,471 COFs.

To validate our MD calculations, we calculated in-plane thermal conductivity $\kappa$ of two 2D COFs with available experimental measurements in the literature \cite{evans2021thermally} (COF-5 and TP-COF; structures presented in \figurename~\ref{fig:MD}d-e) and compared the results in Table~\ref{tab:valid}. For each COF, three replicate simulations with different initial atomic velocities were performed and the mean and standard deviation of $\kappa$ are reported. Our calculations agree reasonably well with experimental results, and the quantitative differences can be attributed to the inherent limitations, such as \textit{(i)} we measured the in-plane thermal conductivity, while the reported values are for the cross-plane properties, and \textit{(ii)} real COFs contain defects, such as vacancies, dislocations and grain boundaries \cite{daliran2024defects, daliran2024probing}. Our simulations consider perfect, defect-free structures and thereby exhibit different $\kappa$ from the experimental measurements.

\begin{table}[h]
    \centering
    \begin{tabular}{c c c c}
        \hline
        {\textbf{COF}} & {\makecell{\textbf{In-plane $\kappa$ (MD Dreiding)} \\ (W/mK)}} & {\makecell{\textbf{In-plane $\kappa$ (MD AIREBO)} \\ (W/mK)}} & {\makecell{\textbf{Cross-plane $\kappa$ (experiment)} \cite{evans2021thermally} \\ (W/mK)}} \\
        \hline
        {COF-5} & {$1.48 \pm 0.04$} & {1.09} & {$1.03$} \\
        {TP-COF} & {$1.06 \pm 0.02$} & {0.74} & {$0.89$} \\
        \hline
    \end{tabular}
    \caption{{In-plane thermal conductivity $\kappa$ of COF-5 and TP-COF calculated by MD simulations in this work and their cross-plane thermal conductivity measured by experiments in the literature \cite{evans2021thermally}.}}
    \label{tab:valid}
\end{table}

{We further conducted additional simulations using the AIREBO potential \cite{stuart2000reactive}, as detailed in Table~\ref{tab:valid}. Consistent with Evans et al. \cite{evans2021thermally}, we applied the carbon atom parameters from the AIREBO force field to all other heavy atoms, including boron, nitrogen, and oxygen, while correctly assigning their atomic masses. Although AIREBO improves the agreement between MD in-plane $\kappa$ and experimental cross-plane $\kappa$ for COF-5, its significantly higher computational cost and the assumption of using carbon parameters for all heavy atoms pose a risk when simulating 2,400 COFs with diverse chemistries and structures. Consequently, the Dreiding force field was selected for generating thermal conductivity data.}

{Although a more quantitative agreement with experimental results would be ideal, such direct comparisons are impractical due to the lack of experimental in-plane thermal conductivity data and the inherent presence of defects like dislocations and grain boundaries in real COF samples. In contrast, our model was trained on idealized monolayer structures. The objective of this work is not to exactly reproduce the experimental cross-plane $\kappa$ through in-plane COF simulations. Instead, we aim to better understand the intrinsic mechanisms of thermal transport in two-dimensional COFs using a deep learning model. Therefore, a direct comparison with experimental values might not accurately reflect the ability of the model to capture the fundamental structure-property relationships.}

To examine the effect of supercell dimension on calculated thermal conductivity, we calculated the average in-plane thermal conductivity (denoted as $\kappa$) of five COFs with thermal conductivity covering the entire range of the dataset and two existing COFs (COF-5 and TP-COF) (\figurename~\ref{fig:MD}b-h) with different supercell dimensions in heat transfer direction, as shown in \figurename~\ref{fig:MD}a. $\kappa$ of all seven COFs converges as the supercell dimension approaches 7 nm, confirming that the chosen dimension is sufficient to minimize the size effect.

\begin{figure}
\centering
\includegraphics[width=1.0\linewidth]{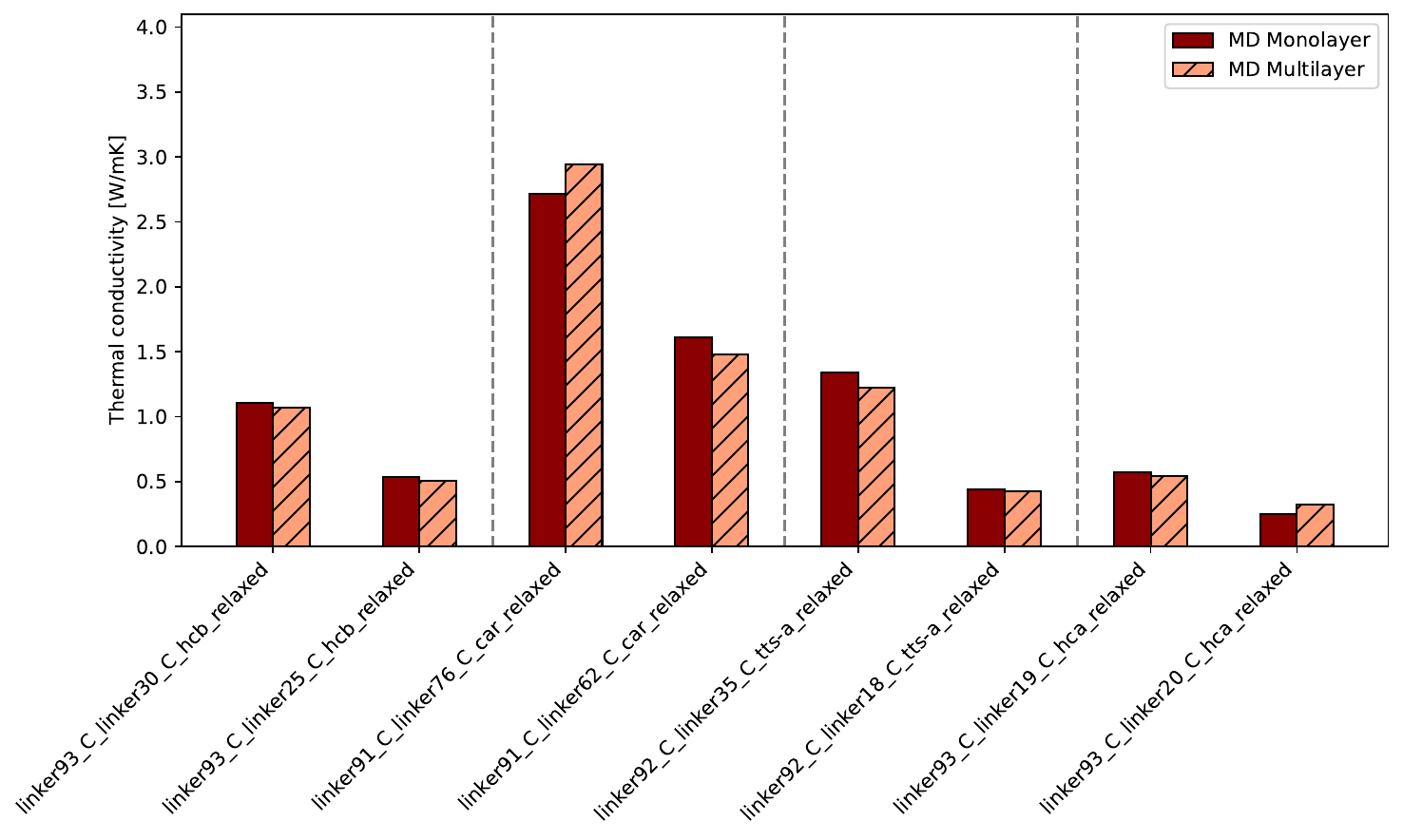}
\caption{{Thermal conductivity values obtained via MD simulations of monolayer and five-layer COFs for the eight selected example COFs  shown in \figurename~4 and 5 in the main article and \figurename~\ref{fig:vdos_SI}.}}
\label{fig:pmt-md-multi}
\end{figure}

{In this work, we simulated monolayer COFs to reduce computational cost. To confirm the effect of cross-plane interactions on calculated thermal conductivity, additional simulations were performed on a selected representative set of eight COFs (\figurename~4 and 5 in the manuscript and \figurename~\ref{fig:vdos_SI}) using five layers. The calculated thermal conductivity for the five-layer COFs closely matches that of the monolayer structures (see \figurename~\ref{fig:pmt-md-multi}), demonstrating a similar effect of dangling atoms in multilayer COF systems.}

The density of each COF was obtained from the equilibrated structure in MD simulations. The structural properties including the largest pore diameter (LPD), void fraction, and gravimetric surface area (GSA) were obtained directly from the dataset by Mercado et al. \cite{mercado2018silico}

\subsection{Minimal anisotropy in in-plane thermal conductivity}

\begin{figure}
\centering
\includegraphics[width=0.7\linewidth]{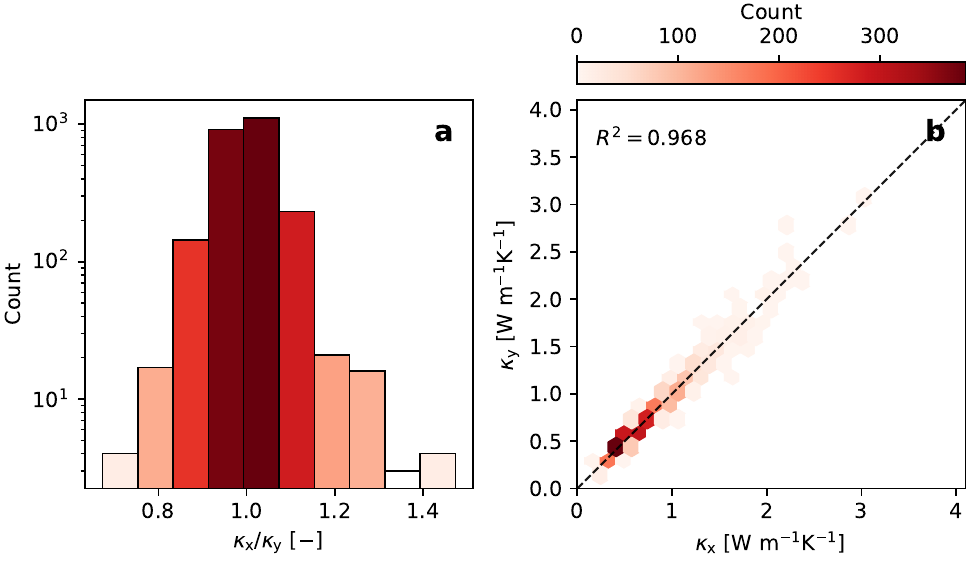}
\caption{ (a) Distribution of the ratio of in-plane thermal conductivities ($ \kappa_{\mathrm{x}}/\kappa_{\mathrm{y}}$), (b) parity plot comparing $\kappa_{\mathrm{x}}$ and $\kappa_{\mathrm{y}}$.}
\label{fig:SI_kxky}
\end{figure}

The distribution in \figurename~\ref{fig:SI_kxky}a shows that the majority of the $\kappa_{\mathrm{x}}/\kappa_{\mathrm{y}}$ values cluster around 1.0, indicating that $\kappa_{\mathrm{x}}$ and $\kappa_{\mathrm{y}}$ are generally similar. This is further supported by the parity plot in \figurename~\ref{fig:SI_kxky}b, which exhibits a strong linear correlation ($\mathrm{R}^2 = 0.968$) between $\kappa_{\mathrm{x}}$ and $\kappa_{\mathrm{y}}$. These observations suggest that for most cases, the in-plane thermal conductivities are nearly equal. Therefore, we utilized average thermal conductivity $\kappa$ as the quantity of interest to explore the structure-property relationships. 

\subsection{{Feature distribution}}
\begin{figure}
\centering
\includegraphics[width=1.\linewidth]{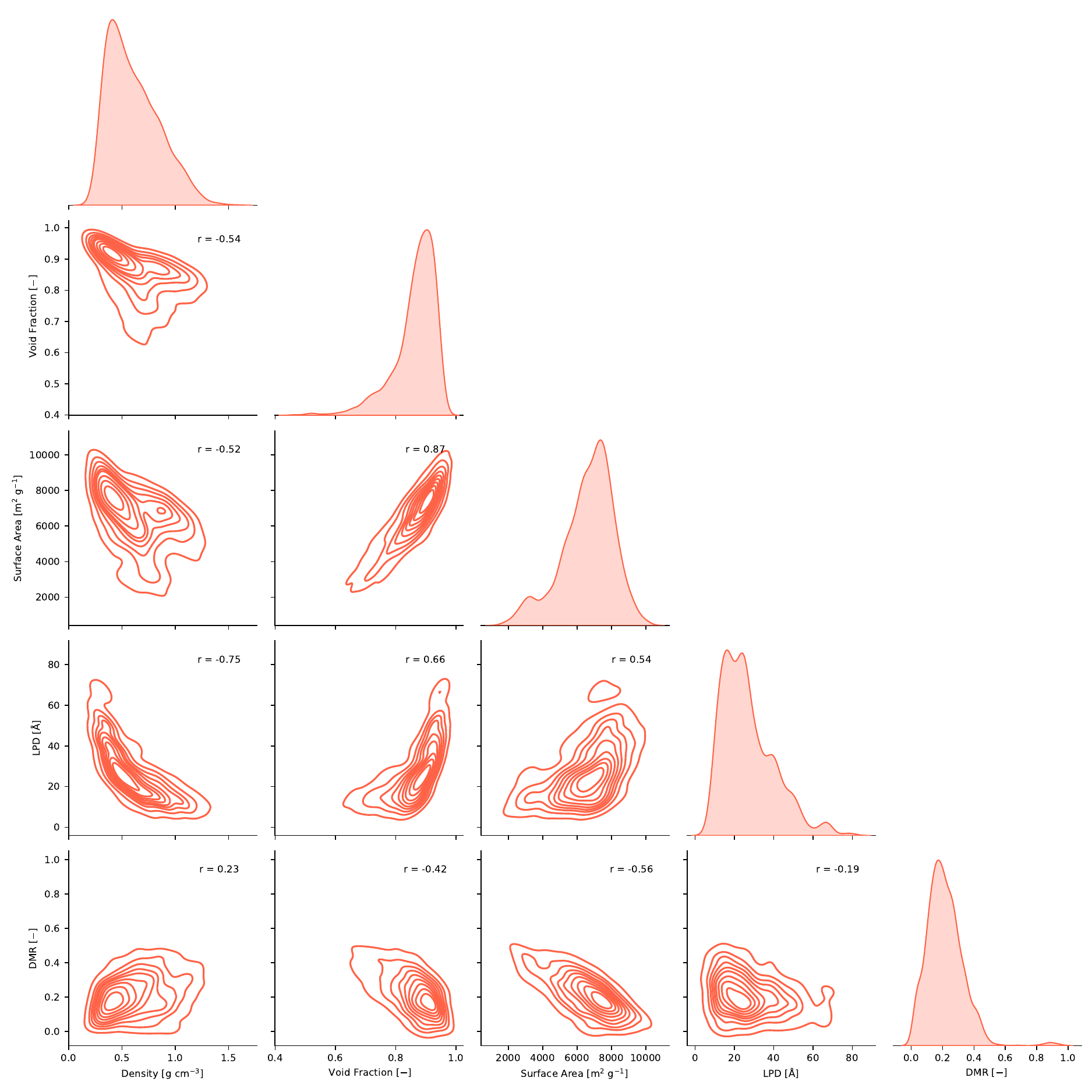}
\caption{{Distribution and pairwise correlations of the five features: density, void fraction, surface area, pore size, and dangling mass ratio. Pearson correlation coefficients are shown in each joint distribution plot.}}
\label{fig:pairplot}
\end{figure}

{\figurename~\ref{fig:pairplot} illustrates the distribution of all features and their intercorrelations. Most features exhibit moderate to strong correlations, including: \textbf{(i)} an inverse correlation between void fraction and density, \textbf{(ii)} an inverse correlation between pore size and density, \textbf{(iii)} a strong positive correlation between void fraction and surface area, and \textbf{(iv)} a positive correlation between pore size and void fraction (and, by extension, with surface area). Notably, the dangling mass ratio (DMR) shows minimal correlation with other features, apart from \textbf{(v)} a moderate inverse correlation with surface area and a weaker inverse correlation with void fraction.}

\begin{figure}
\centering
\includegraphics[width=0.7\linewidth]{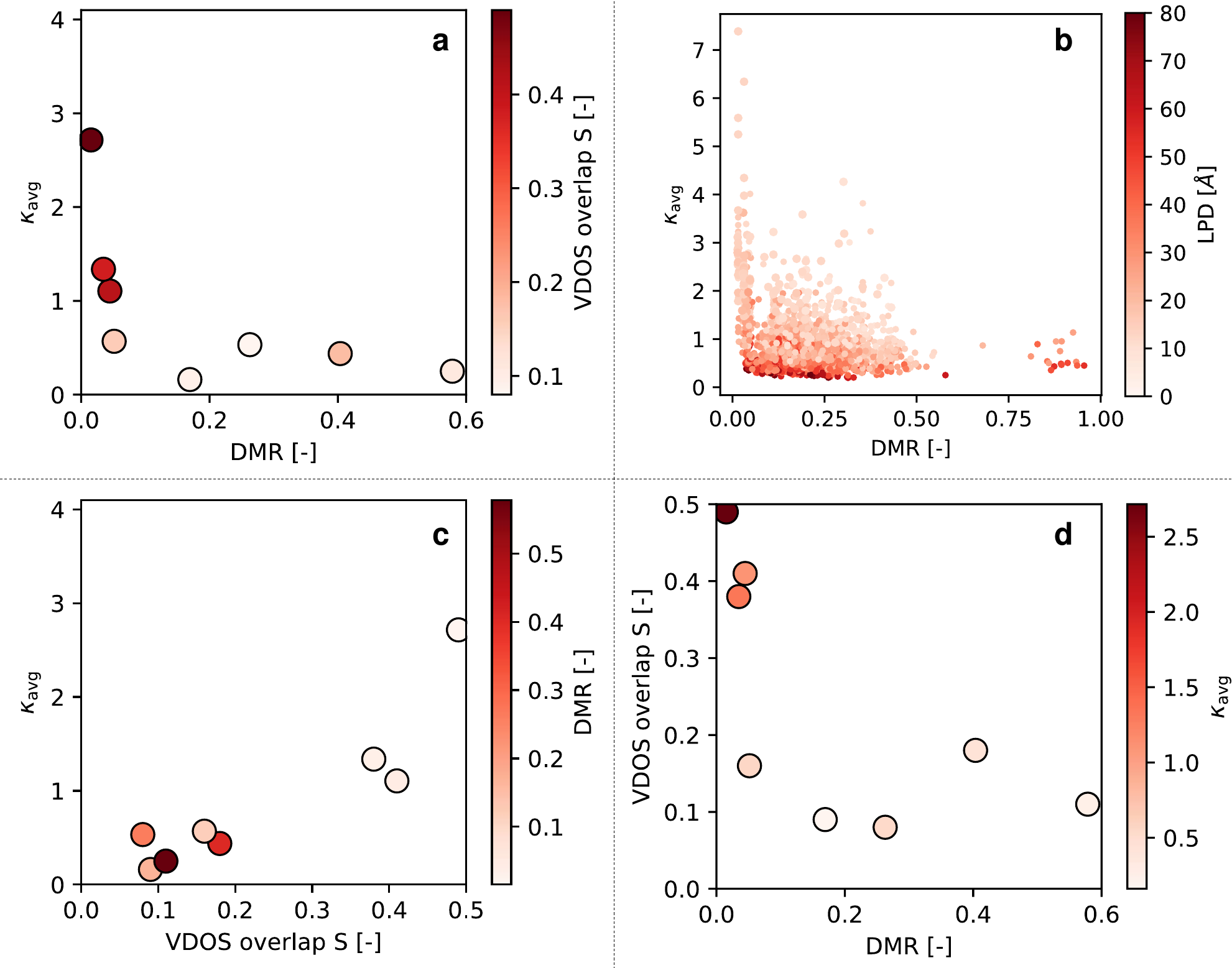}
\caption{{(a) DMR versus $\kappa$ for eight selected COFs presented in \figurename~4 and 5 in the main article and \figurename~\ref{fig:vdos_SI}.  
(b) DMR versus $\kappa$  for the full dataset and additional simulated structures.  
(c) VDOS overlap $S$ versus $\kappa$, and 
(d) DMR versus VDOS overlap $S$ for the same eight selected COFs.}}
\label{fig:DMR-S-k}
\end{figure}

{\figurename~\ref{fig:DMR-S-k}a shows that lower DMR generally corresponds to higher thermal conductivity $\kappa$ for a representative set of eight COFs. \figurename~\ref{fig:DMR-S-k}b extends this analysis to the full dataset and our additional simulations, confirming that COFs with high thermal conductivity tend to have low DMR. Some COFs may exhibit high thermal conductivity despite a high DMR; however, this typically occurs only when other structural features, such as density and pore size, fall within highly favorable ranges. This reinforces DMR as a meaningful, but not exclusive predictor. \figurename~\ref{fig:DMR-S-k}c shows that a higher vibrational density of states (VDOS) overlap $S$ is associated with higher $\kappa$ in the same eight COFs. Additionally, \figurename~\ref{fig:DMR-S-k}d demonstrates that higher DMR tends to reduce VDOS overlap $S$, suggesting that DMR negatively affects phonon coupling.}

\section{{Ensemble Regression Models}}

\begin{table}[h]
\centering
\centering
{
\begin{tabular}{lc}
\toprule
\textbf{Regression Model} & \textbf{$R^2$} \\ 
\midrule
Random Forest              & $\mathbf{0.592\pm0.039}$ \\ 
Gradient Boosting          & $0.536\pm0.038$         \\ 
XGBoost                    & $0.551\pm0.060$         \\ 
AdaBoost                   & $0.312\pm0.103$         \\ 
\bottomrule
\end{tabular}
}
\caption{{Performances of standard ensemble regression models at predicting $\kappa$ using the four standard descriptors (excluding DMR): density, pore size, surface area, and void fraction. The table reports $R^2$ scores obtained through 10-fold cross-validation.}}
\label{tab:regression}

\end{table}

{\textbf{Regression performance excluding DMR as a feature: }We trained four standard ensemble regression models---Random Forest \cite{breiman2001random}, Gradient Boosting \cite{friedman2001greedy}, XGBoost \cite{chen2016xgboost}, and AdaBoost \cite{freund1995desicion}---to predict the thermal conductivities of 2D Covalent Organic Frameworks (COFs) based on four features: Density, Pore Size, Void fraction, and Surface Area. {The predictive performances of all models are outlined in Table~\ref{tab:regression}.}}

{\textbf{Implementation Details:} All models, including those incorporating DMR as a feature, were implemented using the scikit-learn library \cite{scikit-learn}, except for the XGBoost regressor, which was sourced from the xgboost package \cite{chen2016xgboost}. Each model was configured with 100 estimators and trained on a 90/10\% train-test split. To ensure robust evaluation, we performed 10-fold cross-validation to calculate confidence intervals for predictive performance.}

\section{Machine learning details}
We split our dataset randomly into 80\% for training, 10\% for validation and the remaining data for testing. We fine-tune the (pre-trained) PMTransformer and jointly train a prediction head  for 100 epochs. Throughout this fine-tuning phase, the model's performance was assessed at each epoch using the validation subset. This evaluation serves as a checkpoint mechanism to save the model parameters only when there was an observable improvement in the predictive performance on the validation data. This step is necessary in preventing overfitting and ensuring that the model generalizes well to unseen data. Upon completion of the fine-tuning phase, the optimized model was subjected to a final evaluation using the test subset, which had not been exposed to the model during training. The prediction accuracies are reported on the test set. We use a batch size of 32, and fine-tune the model using the AdamW \cite{loshchilov2017decoupled} optimizer at a learning rate of 0.0001, and a default weight decay of 0.01. 

\begin{figure}
\centering
\includegraphics[width=0.7\linewidth]{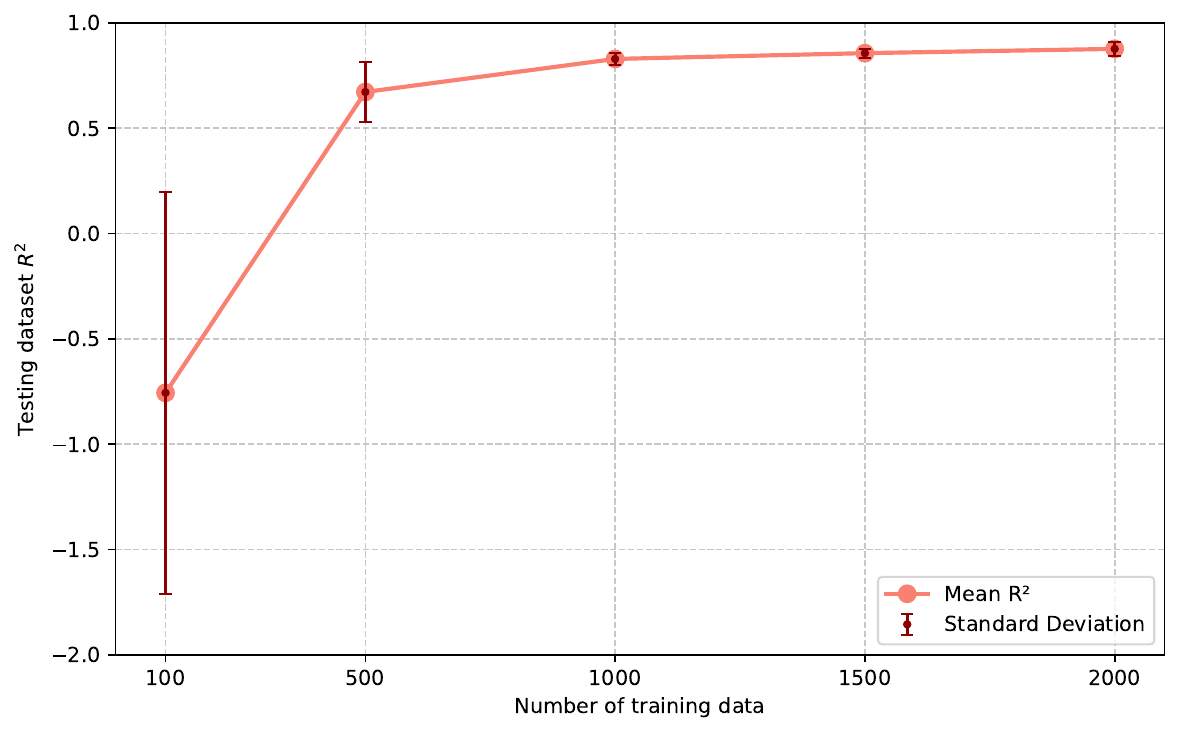}
\caption{{Test $R^2$ scores of the PMTransformer model trained on subsets of 100, 500, 1000, 1500, and 2000 COFs. Each point shows the mean test score, with error bars indicating the standard deviation across five different subsets sampled from the full dataset.}}
\label{fig:ablation}
\end{figure}

{We performed an ablation study to evaluate how the size of the training dataset affects the predictive performance of our model. Specifically, we trained the PMTransformer on five different subsets containing 100, 500, 1000, 1500, and 2000 randomly selected data points, each using an 80/10/10 split for training, validation, and testing. The resulting test $R^2$ scores are reported in \figurename~\ref{fig:ablation}, with error bars representing the standard deviation across five different subsets sampled from the dataset. Test performance saturates at an $R^2$ of 0.8-0.9 when increasing the dataset size to 1000, with minimal standard deviation.}

\begin{figure}
\centering
\includegraphics[width=0.7\linewidth]{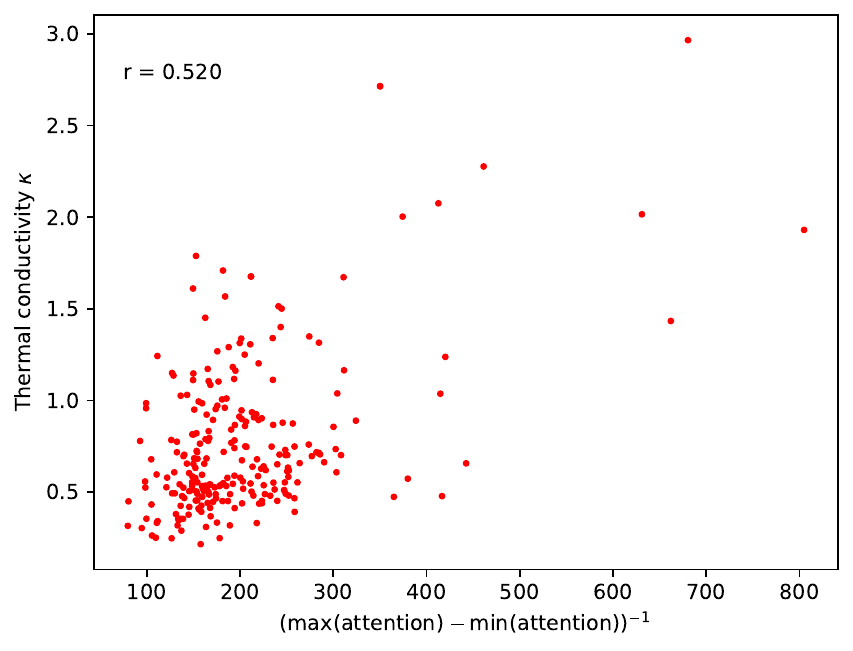}
\caption{\rev{Correlation between thermal conductivity $\kappa$ and the inverse of the attention score range, $(\max(\texttt{attention}) - \min(\texttt{attention}))^{-1}$, for the test set ($N=247$). Here, \texttt{attention} denotes the joint attention across all layers, with each layer’s attention first averaged over all attention heads.}}
\label{fig:attn-k}
\end{figure}

\rev{Figure~\ref{fig:attn-k} illustrates the relationship between the thermal conductivity $\kappa$ and the inverse of the attention score range, $\left( \max(\text{attention}) - \min(\text{attention}) \right)^{-1}$, computed across all atoms in each COF. Here, \texttt{attention} denotes the joint attention across all layers, with each layer’s attention first averaged over all heads. A smaller range of attention scores, corresponding to a more uniform attention distribution, is generally associated with higher thermal conductivity. This trend is observed across the test set ($N=247$), with a Pearson correlation coefficient of $r = 0.52$, which highlights the link between uniform attention profiles and thermal transport properties in COFs.}

\section{Dangling mass computation}

\begin{figure}
\centering
\includegraphics[width=0.7\linewidth]{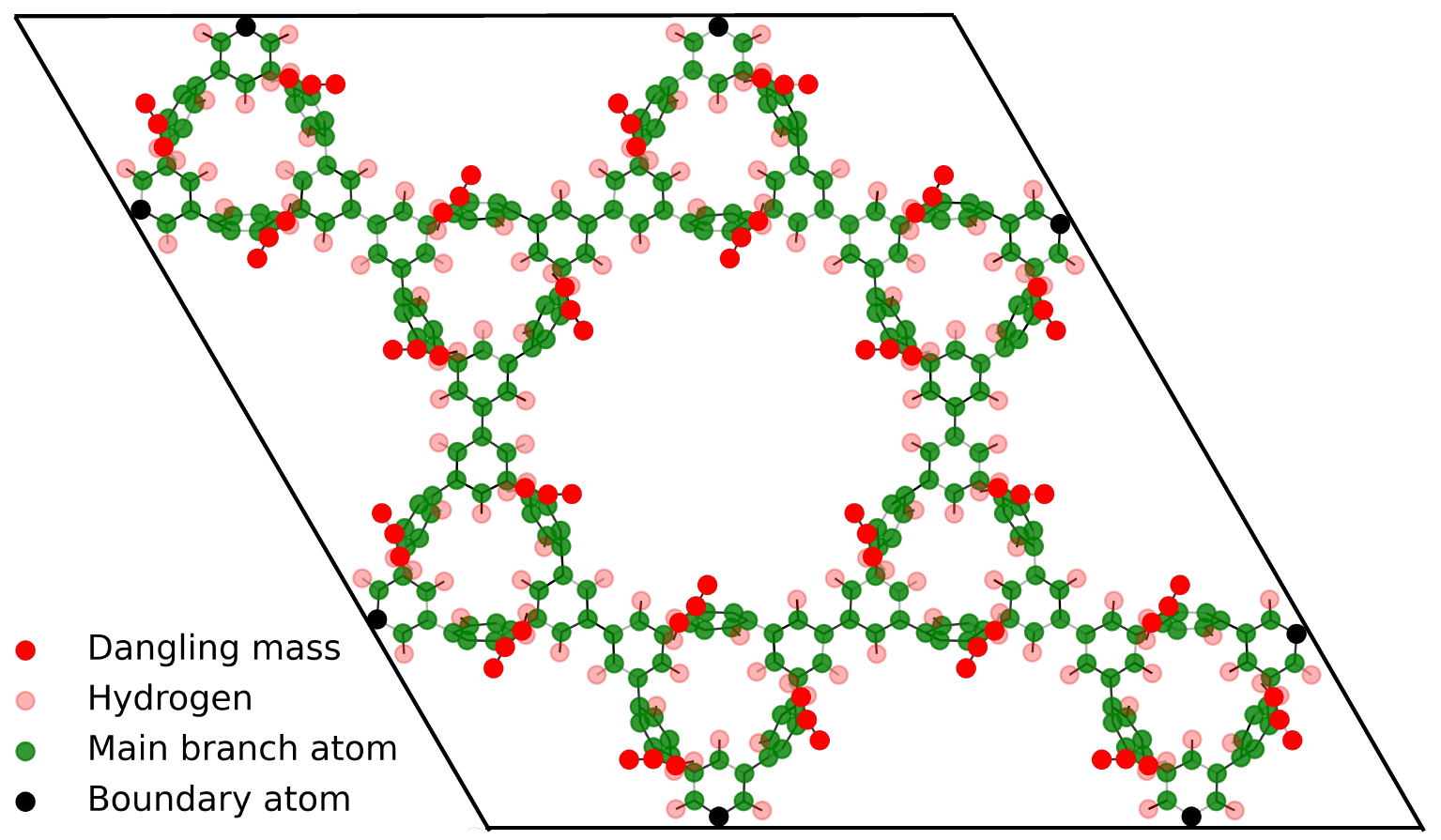}
\caption{A COF unit cell with atoms either labeled as part of the main branch or as dangling atoms. }
\label{fig:SI-dangling}
\end{figure}

Dangling atoms are defined in contrast to the main branch in COF structures. Specifically, within a unit cell, atoms located on the boundary are termed boundary atoms (represented by black atoms in \figurename~\ref{fig:SI-dangling}). The shortest paths that connect all boundary atoms are identified as the main branch (yellow atoms in \figurename~\ref{fig:SI-dangling}). Note that boundary atoms are also labelled as part of the main branch. Atoms that are not part of the main branch are considered dangling branches (red atoms in \figurename~\ref{fig:SI-dangling}). Notably, if a ring contains more than three atoms on the main branch, the ring is considered rigid, and the remaining atoms within the ring are not classified as dangling atoms. The total mass of the dangling atoms in a unit cell is referred to as the dangling mass. To distinguish, dangling hydrogen atoms are shown in lighter red, while the remaining heavier dangling atoms are shown in red.

\section{More examples of COFs and the effect of dangling mass}

\begin{figure}
\centering
\includegraphics[width=\linewidth]{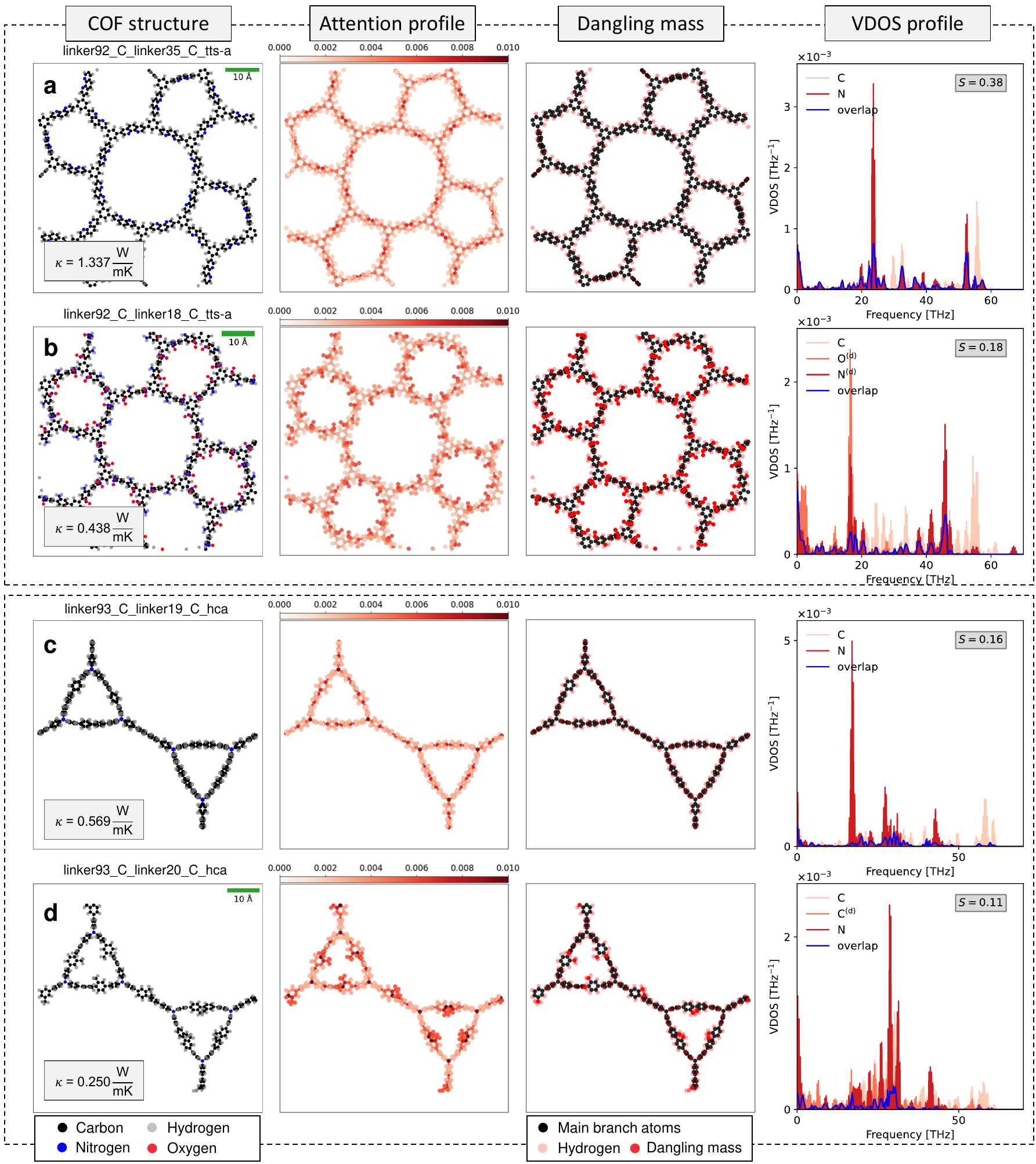}
\caption{Two example pairs (contained in dashed boxes) of COFs with same topologies, similar geometric descriptors, but contrasting  thermal conductivities. The first column illustrates the COF structures. The second column shows the atom-level attention score profile computed by the attention mechanism. The third column shows the same COF structure distinguishing atoms on the main branch and dangling atoms (with separate distinction for hydrogen atoms). The fourth column shows the VDOS profiles of various groups of atoms within the corresponding COF structure with the overlap metric $S$. The legend indicates the VDOS profile for main branch atoms $(.)$ and dangling atoms $(.)^{\text{(d)}}$. {The reported thermal conductivities are obtained from NEMD simulations, rather than predicted by the PMTransformer model.}}
\label{fig:vdos_SI}
\end{figure}

\begin{figure}
\centering
\includegraphics[width=\linewidth]{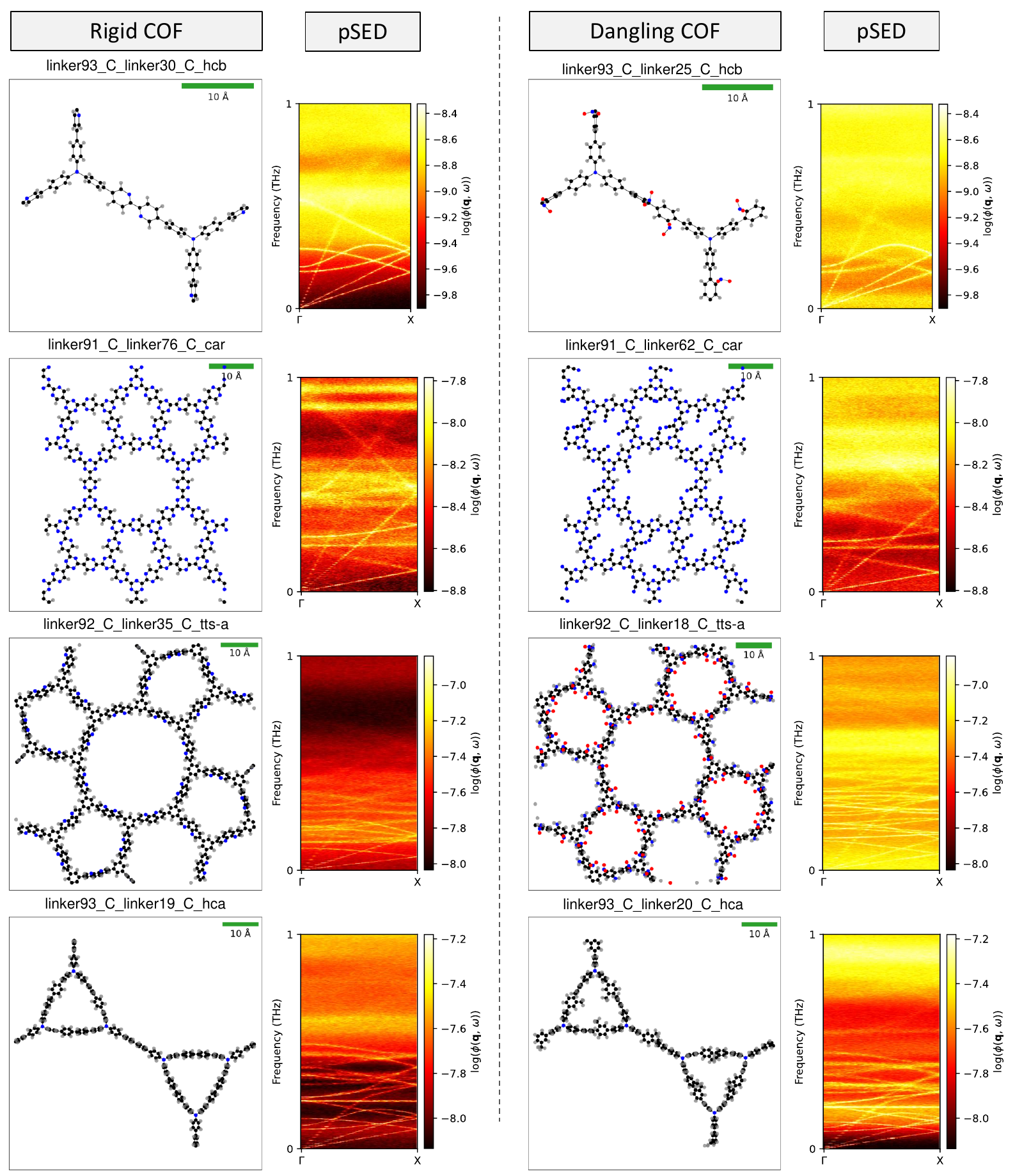}
\caption{{Phonon spectral energy density (pSED) maps of all example COFs {(shown in \figurename~4 and 5 in the main article and \figurename~\ref{fig:vdos_SI})}. For each pair of COFs, the lower bound of the colorbar is set as the minimum value across both pSED profiles. The upper bound is set as the maximum value across both pSED profiles up to 99 percentile to add contrast.}}
\label{fig:psed}
\end{figure}

\begin{figure}
\centering
\includegraphics[width=0.7\linewidth]{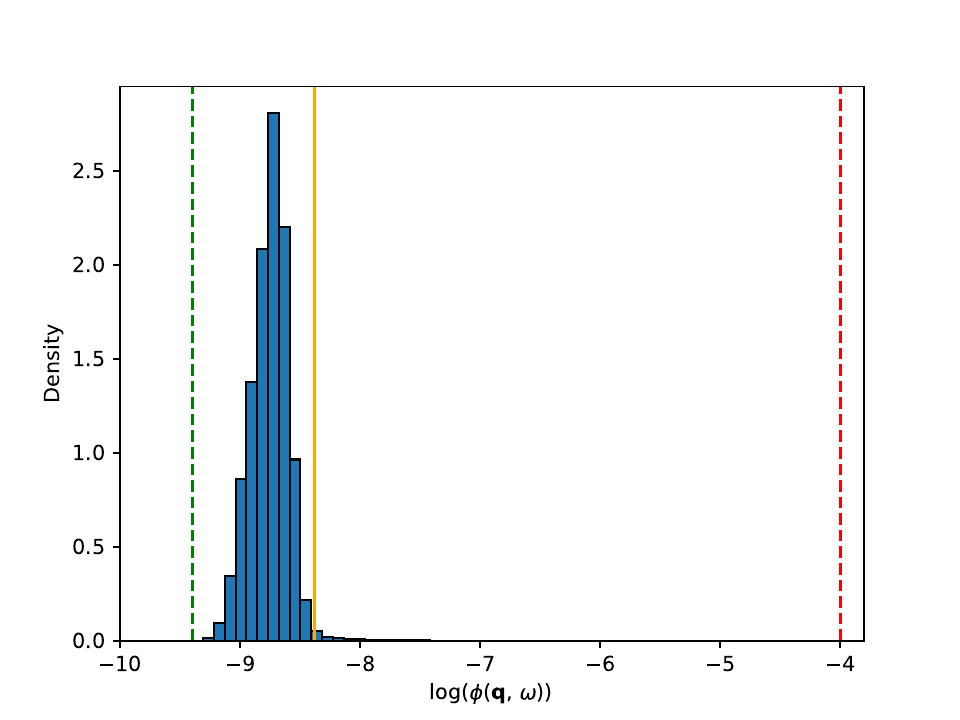}
\caption{{Distribution of logarithm pSED values for an example COF (linker93\_C\_linker25\_C\_hcb, structure presented in \figurename~\ref{fig:psed}).}}
\label{fig:psed_distribution}
\end{figure}

In \figurename~\ref{fig:vdos_SI} we present additional examples of the influence of dangling mass in COFs. In \figurename~\ref{fig:psed} we illustrate the phonon spectral energy density (pSED) maps for all example COFs and their dangling counterparts, which were introduced in \figurename~4 and 5 of the main article and the additional examples in \figurename~\ref{fig:vdos_SI}. For the visualization of the pSED profiles of each COF pair, the lower bound of the colorbar is defined as the minimum value across both logarithmic pSED profiles, while the upper bound is set as the maximum value across both logarithm pSED profiles up to 99 percentile to enhance contrast. These bounds are maintained consistently across each pair to ensure a meaningful comparison. The distribution of logarithmic pSED values for an example COF is presented in \figurename~\ref{fig:psed_distribution}.

\section{Vibrational density of states and phonon spectral energy density calculation}

\begin{figure}
\centering
\includegraphics[width=1.\linewidth]{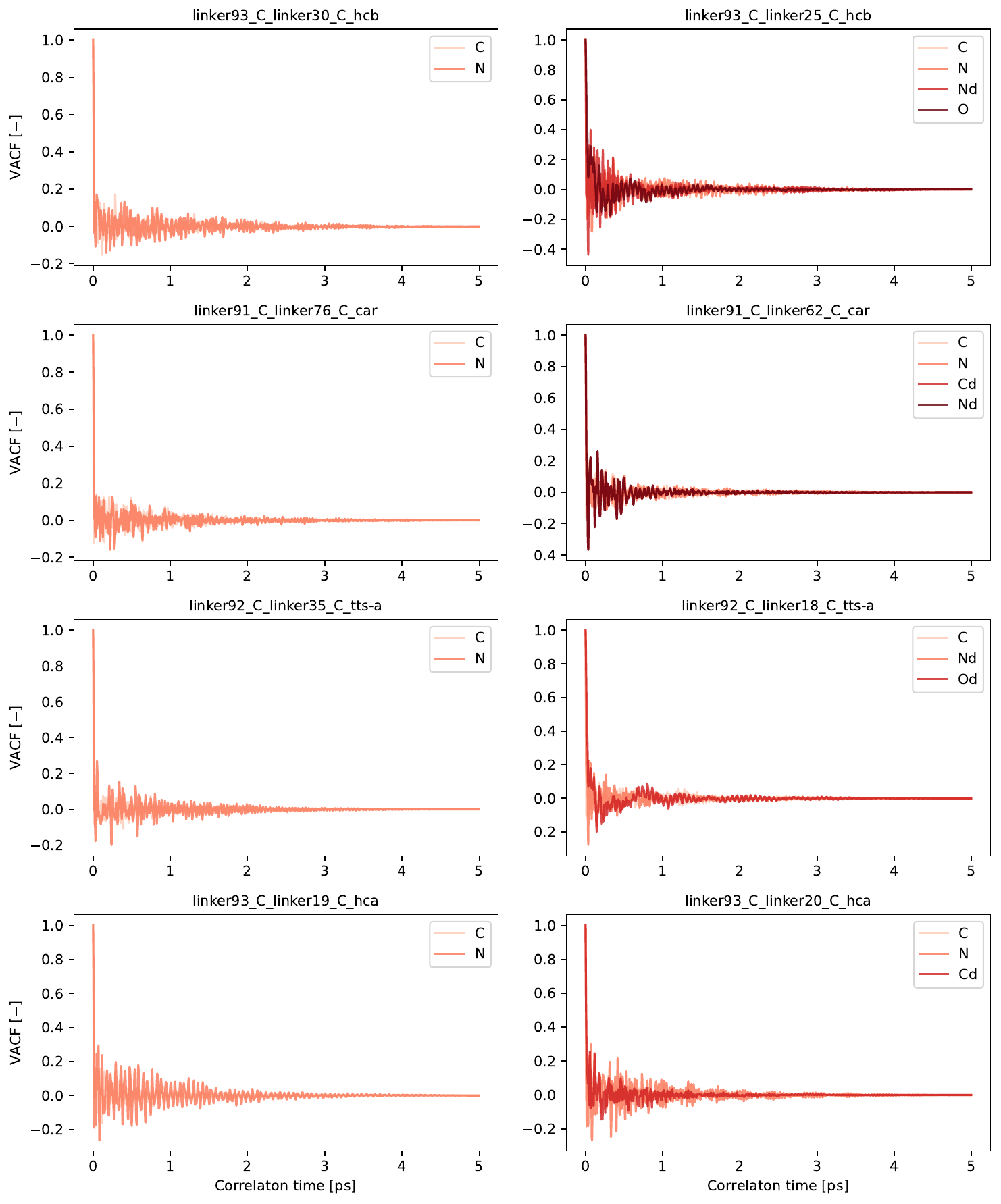}
\caption{{Velocity autocorrelation function (VACF) as a function of correlation time for the eight selected COFs presented in \figurename~4 and 5 in the main article and \figurename~\ref{fig:vdos_SI}.}}
\label{fig:VACF}
\end{figure}

To calculate vibration density of states (VDOS) of atoms, each equilibrated COF structure was first relaxed under NVE ensemble for 1 ns followed by a 0.1-ns NVE production period, where atomic velocities were recorded every 5 fs. VDOS is calculated by the Fourier transform of the normalized velocity autocorrelation function (VACF) of a specific group of atoms $\beta$:
\begin{equation} \label{eq_vdos}
    \mathrm{VDOS}^\beta (\omega) = \mathcal{F} \left( \mathrm{VACF}^\beta (t) \right)
\end{equation}
and
\begin{equation} \label{eq_vacf}
    \mathrm{VACF}^\beta (t) = \frac{\sum_{i \in \{x, y, z\}} \sum_{j \in \beta} \langle v_i^{j}(t_0 + t) v_i^{j}(t_0) \rangle}{\sum_{i \in \{x, y, z\}} \sum_{j \in \beta} \langle v_i^{j}(t_0) v_i^{j}(t_0) \rangle},
\end{equation}
where $\beta$ is the interested group of atoms, $\mathcal{F}$ is the Fourier transform, $\omega$ is frequency, $v_i^{j}$ is the velocity of atom $j$ in $i$ direction, $t$ is correlation time and $\langle \rangle$ is the average over all time origins $t_0$. {The normalized VACF as a function of correlation time is plotted in \figurename~\ref{fig:VACF} which indicates sufficiently long production runs for convergence.}

The phonon spectral energy density (pSED) of COFs was calculated from MD simulations using the following equation \cite{thomas2010predicting}:
\begin{equation}
    \Phi(\mathbf{q}, \omega) = \frac{1}{4\pi \tau_0 N_T} \sum_{\alpha \in \{x, y, z\}} \sum_{b=1}^{B} m_b \left| \int_{0}^{\tau_0} \sum_{n=1}^{N_T} \dot{u}_\alpha(n, b; t) \times \exp\left(i\mathbf{q} \cdot \mathbf{r}(n, 0; t) - i\omega t\right) \, dt \right|^2,
\end{equation}
where $\mathbf{q}$ is the wave vector, $\omega$ is the frequency, $\tau_0$ is the simulation time, $N_T$ is the number of unit cells in the simulation supercell, $\alpha$ is the direction, $b$ is the label of each atom in the unit cell, $B$ is the number of atoms in the unit cell, $m_b$ is the mass of atom $b$, $n$ is the label of each unit cell in the simulation supercell, $\dot{u}_\alpha(n, b; t)$ is velocity in the direction $\alpha$ at time $t$ of atom $b$ in unit cell $n$, and $\mathbf{r}(n, 0; t)$ is the equilibrium position of unit cell $n$. The positions and velocities of each atom were extracted from MD simulations under the NVE ensemble.

\section{Computing time and resource estimates}
\tablename~\ref{tab:compute} lists the runtime and resources used for different tasks. The computing times are meant to provide only a qualitative
impression, as different software architectures and hardware environments were used for each task.
\begin{table}[h]

\centering

\begin{tabular}{|c|c|c|c|}

\hline
Task                                           & Software                & Parallelization \& Hardware & Runtime$^\ddag$      \\ \hline
\begin{tabular}[c]{@{}c@{}}Computing $\kappa$ in one  in-plane\\ direction \end{tabular} &
  LAMMPS \cite{thompson2022lammps}  &
  CPU, no parallelization$^\P$ &
  150 hours$^\dagger$ \\ \hline
  Predicting $\kappa$ with PMTransformer & PMTransformer in Python & 1x NVIDIA A5000 GPU         & 0.07 seconds$^\dagger$ \\ \hline
Energy grid computation                & GRIDAY \cite{githubGitHubSangwon91GRIDAY}                  & CPU, no parallelization     & 3 seconds$^\dagger$    \\ \hline
Fine-tuning the PMTransformer                  & PMTransformer in Python & 3x NVIDIA A5000 GPUs        & 1 hour       \\ \hline
VDOS computation                     & LAMMPS                  & 8 MPI Cores                & 12 hours$^\dagger$          \\ \hline

Training regression models with 10-fold CV     & Scikit-learn in Python  & CPU, no parallelization     & 20 seconds   \\ \hline

\end{tabular}
\caption{Runtime, software, parallelization, and hardware resources used for different tasks. $^\ddag$Runtimes
reported are rough estimates only and may vary across different simulations and hardware. $^\dagger$Runtimes reported for predictions via PMTransformer and MD computations are averages for one sample. $^\P$Tasks are performed on the Oracle Cloud Infrastructure (OCI).} %
\label{tab:compute}
\end{table}

\bibliographystyle{abbrv}
\bibliography{rsc}